\documentclass[bibyear]{aa}

\usepackage{graphicx}
\usepackage[font=small,labelfont=bf]{caption} 
\usepackage{subcaption}
\usepackage{txfonts}
\usepackage{booktabs}
\usepackage{multirow}
\usepackage{tablefootnote}
\usepackage[T1]{fontenc}
\usepackage{booktabs,siunitx}
\usepackage{amsmath}
\usepackage{xcolor}
\usepackage{lipsum}
\newcommand{\orcid}[1]{\href{https://orcid.org/#1}{\includegraphics[width=9pt]{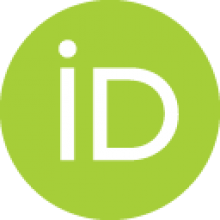}}}
\usepackage{natbib,twoopt}
\usepackage[breaklinks=true]{hyperref}[links=true, link={blue}, cite={blue}]
\hypersetup{
    colorlinks = true,
    linkcolor = {blue},
    citecolor = {blue}
}

\usepackage{xspace}
\newcommand{\AURIGA}{\texttt{AURIGA}\xspace}
\newcommand{\HESTIA}{\texttt{HESTIA}\xspace}
\newcommand{\NIHAO}{\texttt{NIHAO}\xspace}

\definecolor{darkgreen}{rgb}{0.1, 0.6, 0.1}

\begin{document}

\title{A physically motivated galaxy size definition across different state-of-the-art hydrodynamical simulations}

 \author{Elena Arjona-Gálvez \orcid{0000-0002-0462-7519} \inst{1,2}, Salvador Cardona-Barrero \orcid{0000-0002-9990-4055} \inst{2,1}, Robert J. J. Grand \orcid{0000-0001-9667-1340} \inst{3}, Arianna Di Cintio \orcid{0000-0002-9856-1943}
 \inst{2,1}, Claudio Dalla Vecchia \orcid{0000-0002-2620-7056}\inst{1}, Jose A. Benavides\orcid{0000-0003-1896-0424}\inst{4}, Andrea V. Macciò \orcid{0000-0002-8171-6507}\inst{5,6,7}, Noam Libeskind\orcid{0000-0002-6406-0016}\inst{8} \and Alexander Knebe\orcid{0000-0003-4066-8307}\inst{9,10,11}}

\institute{Instituto de Astrofísica de Canarias, Calle Via Láctea s/n, E-38206 La Laguna, Tenerife, Spain\\
\email{eag@iac.es}
\and 
Universidad de La Laguna, Avda. Astrofísico Fco. Sánchez, E-38205 La Laguna, Tenerife, Spain\\
\email{alu0101295794@ull.edu.es}
\and
Astrophysics Research Institute, Liverpool John Moores University, 146 Brownlow Hill, Liverpool, L3 5RF, U
\and
Department of Physics and Astronomy, University of California, Riverside, 900 University Avenue, Riverside, CA 92521, USA
\and
New York University Abu Dhabi, PO Box 129188 Abu Dhabi, United Arab Emirates
\and
Center for Astrophysics and Space Science (CASS), New York University Abu Dhabi
\and
Max-Planck-Institut für Astronomie, Königstuhl 17, D-69117 Heidelberg, Germany
\and
Leibniz-Institut für Astrophysik Potsdam (AIP), An der Sternwarte 16, D-14482 Potsdam, Germany
\and
Departamento de Física Teórica, Módulo 15, Facultad de Ciencias, Universidad Autónoma de Madrid, 28049 Madrid, Spain
\and
Centro de Investigación Avanzada en Física Fundamental (CIAFF), Facultad de Ciencias, Universidad Autónoma de Madrid, 28049 Madrid, Spain
\and
International Centre for Radio Astronomy Research, University of Western Australia, 35 Stirling Highway, Crawley, Western Australia 6009, Australia}

\authorrunning{Arjona-Gálvez E., Cardona-Barrero S., Grand R., Di Cintio, A. \& Dalla Vecchia C.}
\titlerunning{}

   \date{Received XXX XX, XXXX; accepted YYY YY, YYYY}

 
  \abstract
   {Galaxy sizes are a key parameter to distinguish between different galaxy types and morphologies, which in turn reflect distinct formation and assembly histories. Several methods have been proposed to define the boundaries of galaxies, often relying on light concentration or isophotal densities. However, these approaches were often constrained by observational limitations and did not necessarily provide a clear physical boundary for galaxy outskirts.
   }
   {With the advent of modern multiwavelength deep imaging surveys, recent observational studies have introduced a new, physically motivated definition for determining galaxy sizes. This method takes the current or past radial position of the star formation threshold as the size of the galaxy. In practice, a proxy for measuring this position in the present-day universe is the radial position of the stellar mass density contour at 1 M$_\odot$pc$^{-2}$, defined as R$_1$. In this study, we aim to test the validity of this new definition and assess its consistency across different redshifts and galaxy formation models.
   }
   {We analyze three state-of-the-art hydrodynamical simulation suites to explore the proposed size-stellar mass relation. For each simulation suite, we examine the stellar surface density profiles across a wide range of stellar masses and redshifts. We measure the galaxy sizes according to this new definition and compare them with the most traditional size metric, the stellar half-mass radius. 
   }
   {Our analysis demonstrates that the R$_1$-M$_\star$ relation exhibits consistent behaviour across both low and high stellar mass galaxies, with remarkably low scatter. This relation is independent of redshift and holds across the three different cosmological hydrodynamical simulation suites, highlighting its robustness to variations in galaxy formation models. 
   Furthermore, we explore the connection between a galaxy's total mass within R1 and its stellar mass, finding very little scatter in this relation. This suggests that R$_1$ could serve as a reliable observational tracer for a galaxy's dynamical mass.
   }
   {The size-stellar mass relation proposed provides a reliable and physically motivated method for defining the outskirts of galaxies. This method remains consistent not only at $z$=0 but also throughout the evolutionary history of galaxies, offering a robust and meaningful framework for galaxy evolution studies.
   }


\titlerunning{Galaxy size definition across different state-of-the-art hydrodynamical simulations}
\authorrunning{Arjona-Gálvez, E. et al.}

\maketitle

%

%
%
\section{INTRODUCTION}

Measuring the size of galaxies has always been crucial in the astrophysics community, as it strongly correlates with the formation and evolution of these objects \citep{Sersic1968}. However, the unclear boundaries of galaxy outskirts have led to significant discrepancies within the community. Traditionally, two main approaches have been used in the literature. The first approach is based on the radius containing a certain fraction of the galaxy's light. Behind this definition, the most popular way to define the size of a galaxy has been the effective radius, R$_{\rm eff}$, defined as the radial distance containing half a galaxy's total flux \citep{deVaucouleurs1948}. Other examples have also been widely used, such as the radial distance containing 90 per cent of the galaxy's light, R$_{90}$, and the Petrosian or Kron radii \citep{Nair2011,Petrosian1976,Kron1980}. The second approach involves measuring the size of galaxies based on the radial location of fixed surface brightness isophotes. Among these, the most commonly used definitions are $R_{25}$ \citep{Redman1936}, which is based on the isophote at $\mu_B = 25$ mag arcsec$^{-2}$, and the Holmberg radius \citep{Holmberg1958}, $R_H$, defined by the isophote at $\mu_B = 26.4$ mag arcsec$^{-2}$.

Despite the various definition of galaxy size, $\rm R_{\mathrm{eff}}$ has become the most popular among them, and it has even been recently used as a size proxy to investigate the connection between galaxies and halos \citep{Scholz2024}. Observers encounter several limitations in determining the dark matter halo masses of galaxies, being an observational challenge to directly measure them for large samples of galaxies. They are often estimated with alternative methods that have different caveats and limitations (e.g. weak lensing \citep{Mandelbaum2019}, satellite kinematics \citep{More2011} among other quantities that scale with the halo mass). Recently, \cite{Scholz2024} proposed the stellar-to-total dynamical mass relation (STMR) as an alternative metric to link baryons and dark matter within the galaxies, instead of the commonly used stellar mass-halo mass relation (SHMR). The scatter of the SHMR is related to the efficiency of the star formation and, therefore, influenced by a variety of galaxy properties (see \cite{Wechsler2018} and references therein for detailed explanation). On this framework, \cite{Scholz2024} used the total dynamical mass within 3R$_{\rm eff}$, finding that the stellar population properties correlate both with the scatter of the STMR and the one in the SHMR, in a very similar manner, with the total dynamical mass also correlating with the halo mass at fixed stellar mass.

However, $\rm R_{eff}$ is strongly dependent on the shape of the galaxy's light distribution and does not describe the global size of galaxies. Like other methods for defining the size of a galaxy, R$_{\rm eff}$ was initially introduced for operational purposes, driven by the typical depth of optical images available at the time, and was not intended to convey any physical meaning nor describe the full extent of galaxies \citep{Graham2019, Trujillo2001}. \cite{Graham2019} pointed out that such rather arbitrary effective parameters need to be carefully considered, and explore a range of alternative radii, including the position at which the projected intensity drops by a fixed percentage. Despite these limitations, such size definitions have been used in the calibration of state-of-the-art hydrodynamical simulations, such as \texttt{EAGLE} \citep{Crain2015} and \texttt{IllustrisTNG} \citep{Pillepich2018} as well as in significant theoretical studies on the connection between size and internal galaxy properties (see \cite{Chamba2020Review} and the references therein, for a comprehensive review on galaxy size measurements). Oppositely, those definitions that did have a physical basis were often difficult to reproduce by different authors due to their complexity (e.g., the scale length $r_d$ size definition \citep{Mo1998,Mo2010}). To address this issue, \cite{Trujillo2020} proposed a new physically motivated definition, taking advantage of the new deep imaging surveys era. They define the size of a galaxy as the farthest radial location where the gas has enough density to collapse and form stars.
The critical gas surface density threshold for star formation is estimated to be between 3 and 10 M$_\odot$pc$^{-2}$ \citep{Schaye2004}. \cite{Trujillo2020} used a stellar mass density isocontour of 1 M$_\odot$pc$^{-2}$ as a proxy for this critical gas density threshold required for the star formation \citep{Martinez-Lombilla2019}, referring to the radial location of this isodensity contour as $\rm R_1$. Their study found a size-stellar mass relation with an intrinsic dispersion of approximately 0.06 dex for galaxies with M$_\star>10^7$M$_\odot$, which is about three times smaller than the dispersion observed when using the effective radius ($\sim$ 0.15 dex). \cite{SanchezAlmeida20} examined the reason for this reduced scatter, demonstrating that any two galaxies with the same stellar mass share at least one radius with identical surface density, thereby leading to a more consistent size measurement.

The above galaxy size definition can be easily reproduced not only for other deep galaxy observations but also tested across different state-of-the-art simulations. However, while the proxy value for the surface density threshold of $1$ M$_\odot$pc$^{-2}$ has been adopted to define the edges of Milky-Way-like galaxies \citep{Martinez-Lombilla2019}, the question of whether this proxy value is appropriate to define the edges of galaxies of different stellar mass and morphology arises.
In this context, \cite{Chamba2022} examined the star formation threshold by identifying changes in slope or truncation in the radial stellar mass density profiles of different types of galaxies. They define the edge of a galaxy, $\rm R_{edge}$, as the outermost radial location where a significant drop occurs, finding an average stellar mass surface density threshold of $\sim$$3 \rm M_\odot pc^{-2}$ for elliptical, $\sim$$1 \rm M_\odot pc^{-2}$ for spiral and $\sim$$0.6 \rm M_\odot pc^{-2}$ for dwarf galaxies. Their results suggest that while $\rm R_{edge}$ is, by definition, more accurate, it does not have a major impact on the structure of the size-stellar mass relation to that using $\rm R_1$ at $z$=0.

Following the methodology prescribed by \cite{Chamba2022} to estimate R$_{\rm edge}$, \cite{Buitrago2024} studied the size evolution of MW-like disc galaxies, finding that galaxies with a stellar mass of approximately $\sim$5$\times$$\rm10^{10} M_\odot$ have doubled in size since $z$=1. This finding contrasts with previous studies that used the effective radius as a size indicator (see \cite{Nedkova2021,Kawinwanichakij2021} and references therein), which did not observe a significant increase in size as they evolved. Although these two results are not contradictory, as they reflect the different nature of the size measurements used, other observational studies using R$_{\rm eff}$ also found a dependence between galaxy sizes and redshift \citep[see for example][]{Wel2014}. How $\rm R_1$ evolves with time remains an open question, which is compounded by the fact that observations are unable to track the size evolution of individual/the same sample of galaxies over time. To shed light on these aspects of galaxy evolution, we must turn to cosmological simulations.

In a companion work, \cite{Claudio2025} inspect the implications of using a specific gas density threshold to measure galaxy sizes in two simulated cosmological volumes of different resolutions from the \texttt{EAGLE} simulation project \citep{Crain2015}, calibrated to give similar global relations at $z$=0.
In this paper, we explore this same size definition for three different suites of state-of-the-art hydrodynamical cosmological simulations in different environments. The motivation for examining several hydrodynamical simulation suites arises from the fact that the different baryonic processes implemented in each of them, such as star formation and feedback, strongly influence the in-situ evolution of the galaxy and, consequently, their extension. By selecting different suites from different galaxy formation models we aim to assess how variations in these models impact galaxy formation and how this is reflected in the size-stellar mass relation. In section \ref{sec:sample}, we describe the zoom-in simulations used in this work: \NIHAO \citep{Wang2015}, \AURIGA \citep{Grand2017} and \HESTIA \cite{Libeskind2020}. We explain the method to compute the physical motivated parameter used in \cite{Trujillo2020} for each simulated galaxy with stellar masses between 10$^6$ M$_\odot$ < M$_\star$ < 10$^{12}$ M$_\odot$ in section \ref{sec:method}. We show that a tight correlation exists between $\rm R_1$ and the stellar mass of galaxies, for the same stellar mass range studied in observations (section \ref{sec:R1}). To investigate the evolution of R$_1$ versus M$_\star$ with redshift, we examine the relation from $z$$\sim$1 to the present in section \ref{sec:redshift}, finding that the scatter in the relation decreases at higher redshifts. Finally, the dependence of this new galaxy size definition on halo properties and morphology are discussed in sections \ref{sec:halo} and \ref{sec:morpho}. Our main conclusions are summarised in section \ref{sec:conclusions}.

\begin{table}
	\centering
  	\caption{Resolution parameters and number of total galaxies retained in the three different simulation suites at $z$=0.
  }
	\begin{tabular}{|c|c|c|c|}

    &\AURIGA & \HESTIA & \NIHAO \\\hline
    \midrule
\multicolumn{4}{c}{Resolution parameters}
\\
\midrule
\hline
$m_{dm}$ [M$_\odot$]&3$\times$10$^{5}$&1.2$\times$10$^6$&1.7$\times$10$^6$-3.4$\times$10$^3$\\\hline
$m_{gas}$ [M$_\odot$]&5$\times$10$^4$&1.8$\times$10$^5$&3.2$\times$10$^5$-6.2$\times$10$^2$\\\hline
$\epsilon_{z=0}$ [pc]&369&340&931.4-116.4\\\hline
    \midrule
\multicolumn{4}{c}{Selection criteria}
\\
\midrule
\hline
    \small{$\textrm{fMhires} > 0.99$} & \multirow{2}{*}{\texttt{630}} &  \multirow{2}{*}{\texttt{402}} & \multirow{2}{*}{\texttt{162}} \\
    \small{$\textrm{Nstars} > 100$} &    &   &   \\\hline

    \end{tabular}
 \\
\tablefoot{The selection criteria used in each case can be found in the bottom-left box. Host galaxies were selected based on a mass fraction of dark matter in high-resolution particles (fMhires) and a minimum number of star particles (Nstars) for each halo. For \NIHAO simulation suites, we only show the maximum and minimum values of the resolution parameters, as well as the dark matter softening length.}
 \label{tab:criteria}
\end{table}

\section{DATA AND SAMPLE SELECTION}
\label{sec:sample}
This work utilizes state-of-the-art hydrodynamical simulations, which are widely used for comparison to observational studies. Specifically, we employ cosmological zoom-in simulations; \AURIGA and \NIHAO, as well as full local group simulations - \HESTIA- to cover a broad range of masses, resolutions and environments. The resolution parameters for the three simulations are detailed in Table \ref{tab:criteria}. Both the \AURIGA and \HESTIA simulations are run with \texttt{AREPO} \citep{Springel2010, Pakmor2016}, a parallel N-body, second-order accurate magnetohydrodynamics (MHD) code. In addition, \NIHAO is based on the N-body SPH solver \texttt{GASOLINE2} \citep{Wadsley2017}. Both \texttt{AREPO} and \texttt{GASOLINE2} use the cosmological parameters obtained by \cite{Planck2014}: $\Omega _m$=$0.307$, $\Omega _b$=$0.048$, $\Omega _{\Lambda}$=$0.693$, $\sigma_8$=$0.8288$, and a Hubble constant of $H_0$=$100\,h\,\rm km\,s^{-1}\,\rm Mpc^{-1}$, where $h$=$0.6777$.

The three simulation suites in this study were chosen due to their different galaxy evolution models and environments. The \AURIGA sample includes 30 isolated halos\footnote{These simulations correspond to the \texttt{Original}/4 set of publicly available simulations described in \citet{Grand2024}}, with an halo mass range between 9$\times$10$^{11}$ and 2$\times$10$^{12}$ M$_\odot$, selected to represent a mass range that encompasses the halo mass of the Milky Way \citep{Fritz2020}, however recent estimates have extended the MW halo mass range to lower values, i.e. $\sim$1.81$^{+0.06}_{-0.05}$$\times$10$^{11}$M$_\odot$ \citep{Xiaowei2024}. 
Our HESTIA sample consists of the three highest resolution simulations of the Local Group (LG), run within the CLUES framework \citep{Gottloeber2010} using initial conditions constrained by the cosmic flows-2 \citep{Tully2013} catalogue of peculiar velocities (see \cite{Carlesi2016} and \cite{Libeskind2020} for a more in-depth description) and employing the \AURIGA galaxy formation model \citep{Grand2017}. This allows us to analyse the same physics model for both isolated MW-like galaxies and their surrounding galaxies as well as those in a LG environment. The \NIHAO project includes a sample of approximately 100 isolated hydrodynamical cosmological zoom-in simulations, covering a large range of halo masses, from 2.8$\times$10$^{12}$M$_\odot$ to 3.5$\times$10$^{9}$M$_\odot$. Each simulation sample employs different feedback schemes, which play a crucial role in regulating the size and evolution of their galaxies. For more details on the feedback processes used in each sample, we refer readers to \cite{Grand2017} (\AURIGA), \cite{Libeskind2020} (\HESTIA) and \cite{Wang2015} (\NIHAO).

For each simulation, halos and subhalos are identified using the Amiga Halo Finder, \texttt{AHF} \citep[][]{Knollmann2009} following the same standard parameters across all simulations to identify the halos. The halo masses, M$_{200}$, are defined as the mass contained within a sphere of radius R$_{200}$, which contains a density of $\Delta_{200}$$\simeq$200 times the critical density of the Universe at $z$=$0$. Central halos, also known as host halos, are identified as those with the minimum gravitational potential in the group, while all other subhalos within the same group are classified as satellites of the host. Subhalos are considered resolved if they contain at least 200 particles. The primary analysis is conducted using a modified version of \texttt{PYNBODY} \citep[][]{Pontzen2013} which is compatible with \AURIGA, \HESTIA and \NIHAO.

\begin{figure}[t!]
\centering
\includegraphics[width=\columnwidth]{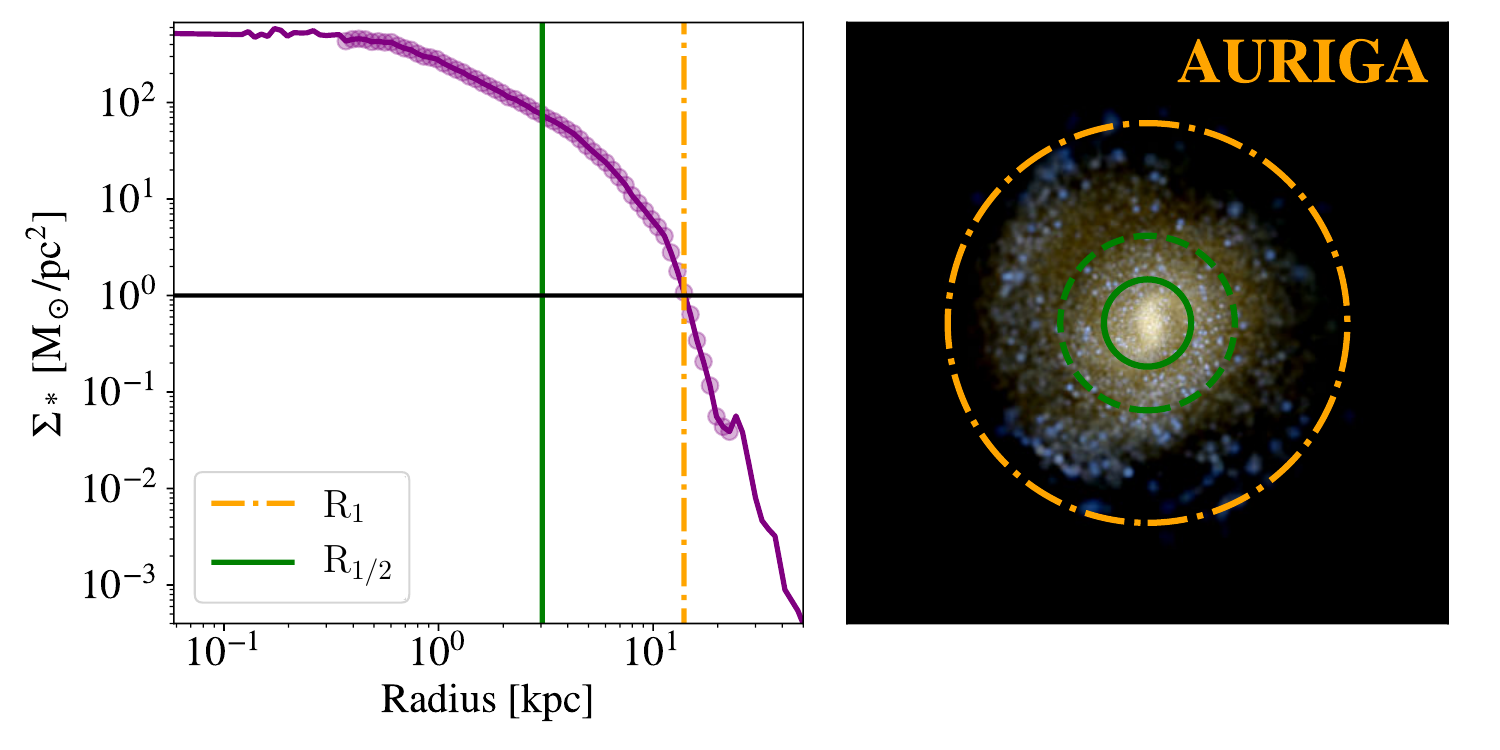}
\vspace{-0.35cm}
\vspace{-0.5cm}
\vspace{-0.45cm}
\includegraphics[width=\columnwidth]{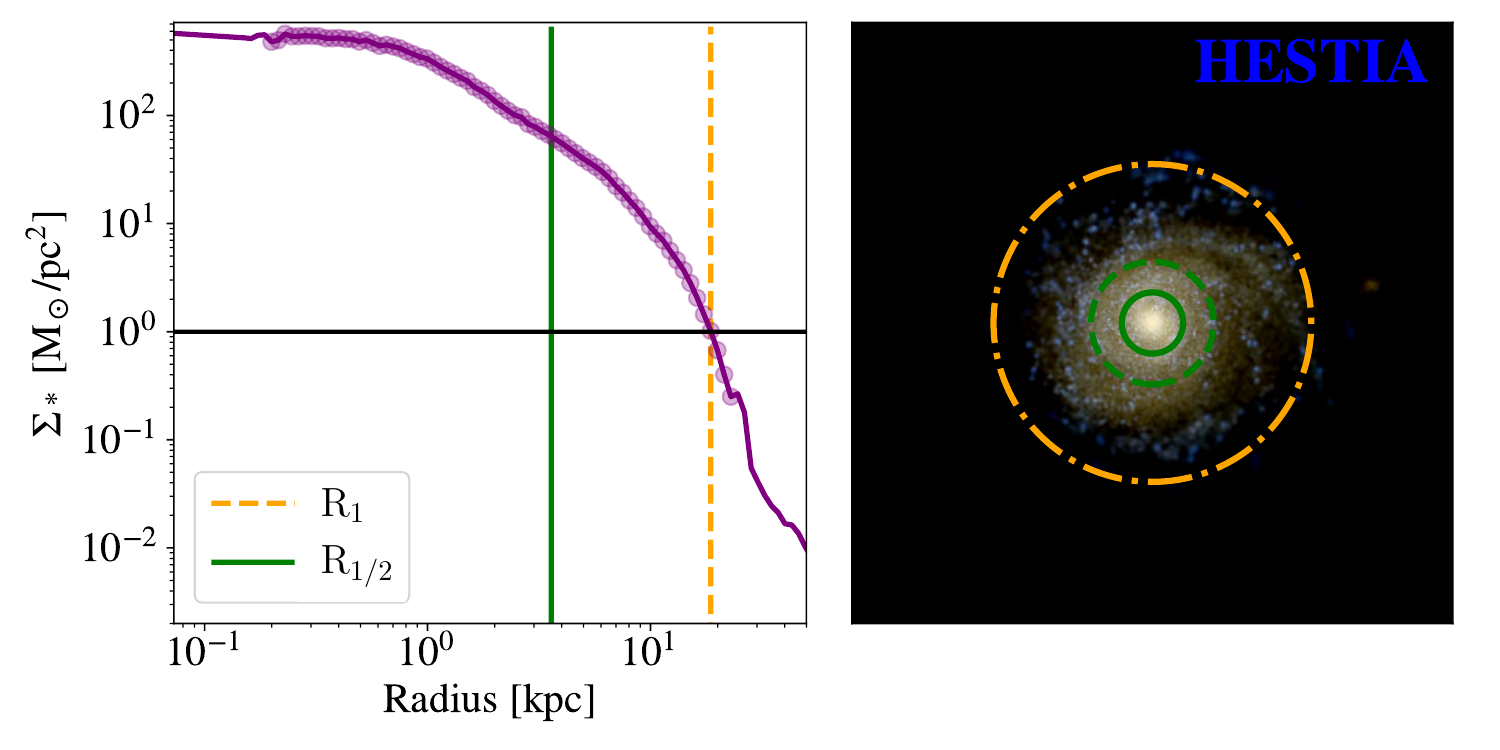}
\includegraphics[width=\columnwidth]{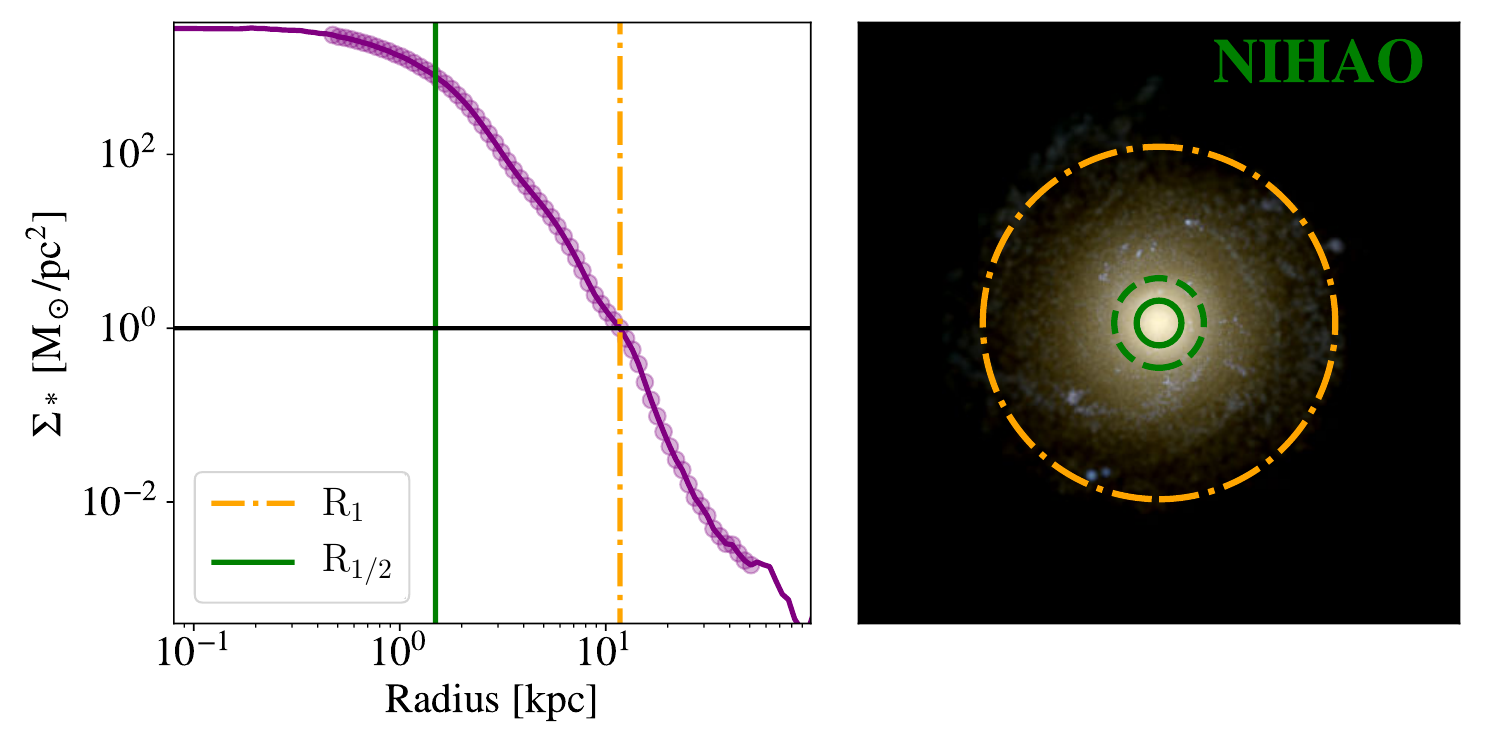}

\caption{Left-hand panels: Stellar surface density profile for three MW-like galaxies in the sample at $z$=0. Right-hand panels: face-on projected stellar light images of the corresponding galaxies. colours are based on the i-, g- and u- band luminosity of stars.  As indicated on the right-hand panels, every row refers to a simulation suite. From top to bottom: \AURIGA, \HESTIA and \NIHAO with the corresponding stellar masses of 1.07$\times$10$^{10}$M$_\odot$, 1.56$\times$10$^{10}$M$_\odot$ and 2.15$\times$10$^{10}$M$_\odot$. Green solid and orange dot-dashed lines indicate the stellar half-mass and R$_1$ radius, respectively. On the right-hand panels, a green dashed line is also incorporated to show the position of twice the stellar half-mass radius.}
\label{fig:fig1}
\end{figure}

\begin{figure*}[h!]
    \centering
    \includegraphics[width=\textwidth]{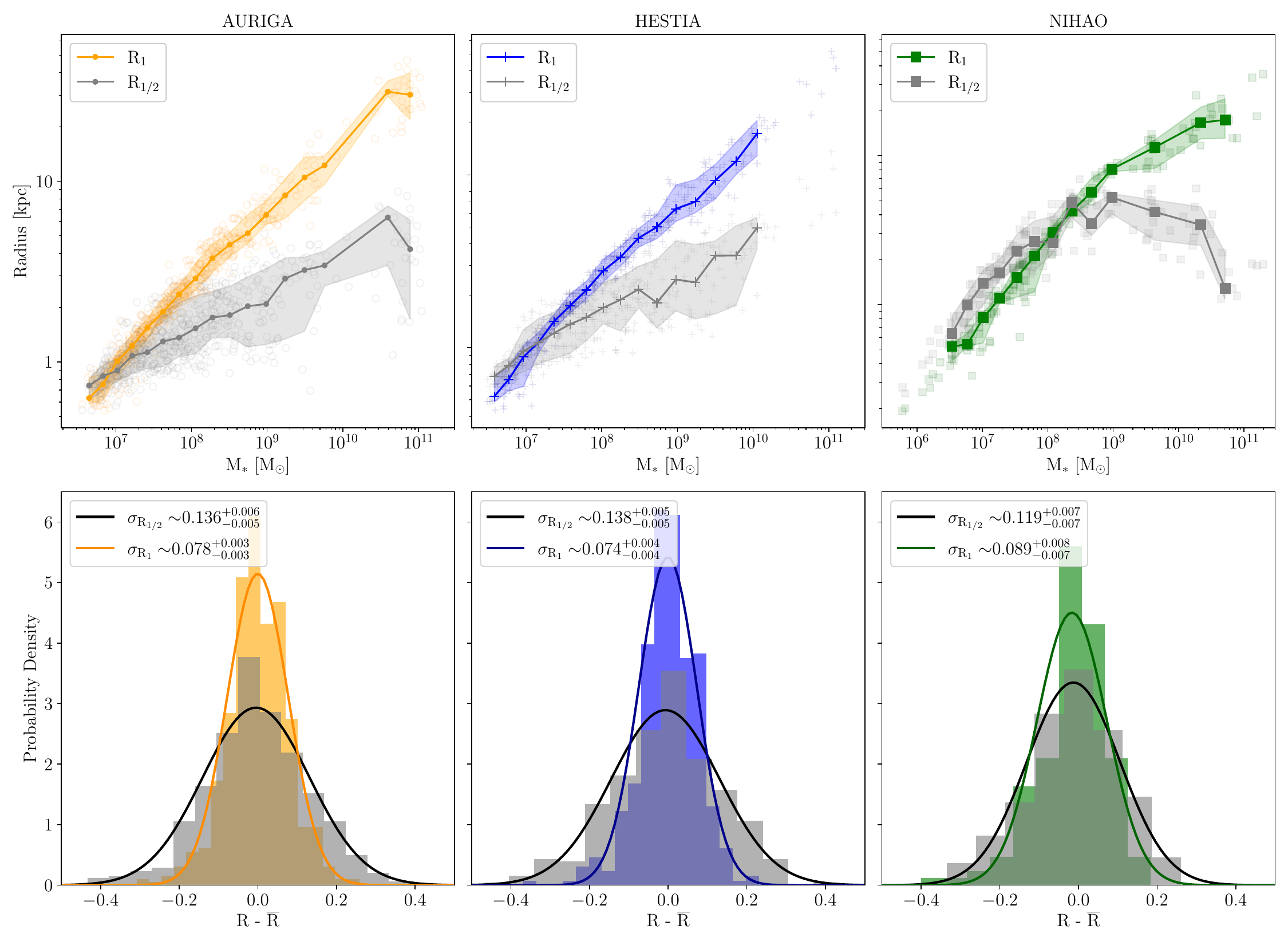}
    \caption{Top panels: size-stellar mass relation for each of the simulation suites used in this work, where the stellar mass is measured within 20\% of R$_{\rm 200}$. The median values and 1-$\sigma$ scatter of the stellar half-mass radius in each stellar mass bin are shown by the grey lines and shaded regions, respectively. Likewise, the median values and 1-$\sigma$ scatter of galaxy size defined as R$_1$ is shown for \AURIGA (orange), \HESTIA (blue) and \NIHAO (green) by the lines and shaded regions, respectively.  The sizes of individual galaxies are shown with dots (\AURIGA), crosses (\HESTIA) and squares (\NIHAO), using the same colour criteria. Bottom panels: normalized histograms of the distribution of galaxy sizes relative to the median ($\overline{\rm R}$) for each simulation suite. In each case, the global $\sigma$ of the relation is found by fitting a Gaussian distribution (solid line following the same colour of the respective histogram) and $\sigma$ is defined as the FWHM of the respective Gaussian, whose errors are computed via bootstrapping.}

    \label{fig:comparison}
\end{figure*}

\begin{figure}[h!]
\hspace{-1.5cm}
\includegraphics[width=1.4\columnwidth]{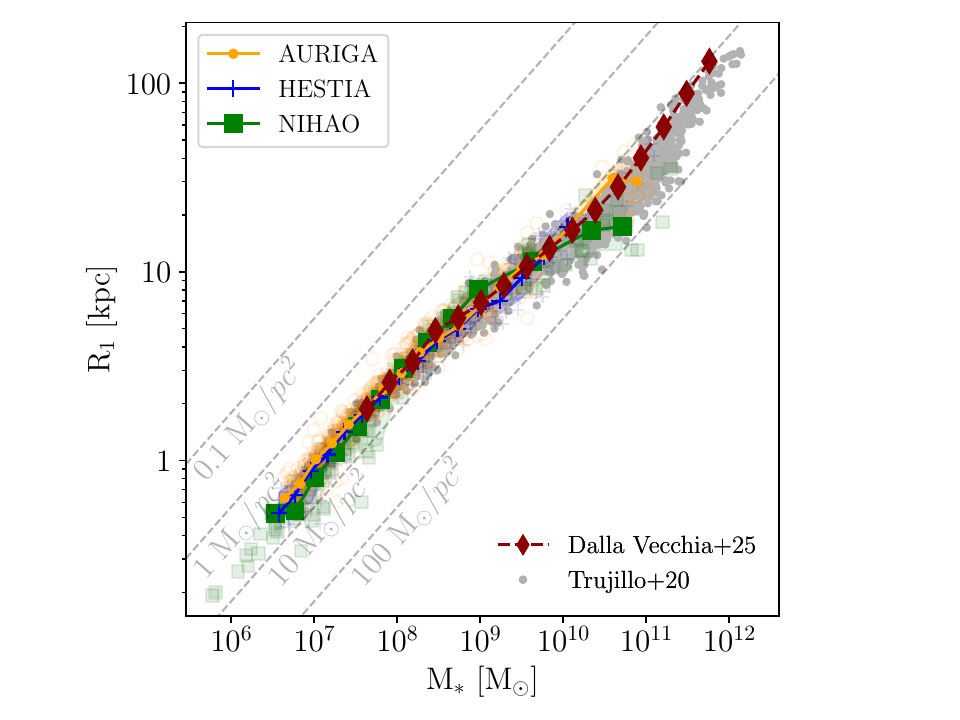}
   \caption{R$_1$ as a function of the stellar mass of galaxies, measured within 20\% of R$_{\rm 200}$, for different simulation suites as indicated in the legend. We show the median value of R$_1$ measured within 20 stellar mass bins, equally spaced. Orange, blue and green relations correspond to the \AURIGA, \HESTIA and \NIHAO sample. In addition, we show the relation found by \cite{Claudio2025}, using public data from \texttt{EAGLE} 
   simulation suite in red. To compare with \cite{Trujillo2020}, we add in grey their observational results as well as dashed lines, corresponding to locations in the place with a constant projected stellar mass density of (from top to bottom) 0.1, 1, 10 and 100 M$_\odot$/pc$^{2}$.}
   
   \label{fig:R1}
\end{figure}

To ensure a fair comparison between samples, we applied the same selection criteria to all zoom-in simulations: galaxies are selected if they contain fewer than $1\%$ low-resolution particles and at least 100 star particles (see table \ref{tab:criteria}). We selected 1251 central galaxies for \AURIGA, 710 for \HESTIA and 350 for \NIHAO using these criteria. Of these, we retain only those galaxies that have a stellar surface density value of 1 M$_\odot$/pc$^2$ at a radius beyond the physical softening length (see \cite{Grand2017,Libeskind2020,Wang2015} for more details about the physical softening length for each simulation), which gives final samples of 630 galaxies for \AURIGA, 402 for \HESTIA and 162 for \NIHAO, with stellar mass ranges of (3.6$\times$$10^6$, 1.1$\times$$10^{11}$), (3.1$\times$$10^6$, 1.3$\times$$10^{11}$) and (4.9$\times$$10^5$, 1.9$\times$$10^{11}$) M$_\odot$ for galaxies at $z$=0, respectively (see Table \ref{tab:criteria}).

\section{METHODOLOGY}\label{sec:method}

In this paper, we explore a new definition of galaxy size introduced by \cite{Trujillo2020}, and apply it to several zoom-in hydrodynamical simulations. Each simulation has used the Amiga Halo Finder code, \texttt{AHF} \citep{Knollmann2009}, to identify halos and subhalos inside the simulations. We selected galaxies at $z$=0 following the selection criteria (see section \ref{sec:sample} and table \ref{tab:criteria} for further discussion) and also examined galaxy evolution by applying the same criteria at $z$$\sim$ 0.2, 0.5, 0.7 and 1. Additionally, we tracked the size evolution of $z$=0 galaxies across four different stellar mass ranges- 10$^7$, 10$^8$, 10$^9$ and 10$^{10}$ M$_\odot$- throughout their evolutionary history.

For all galaxies, we measure the face-on\footnote{The face-on projection is defined as the plane whose normal is the total angular momentum of stars within 20 $\%$ of R$_{200}$ } 2D stellar mass-density profile. We remove any "peaks" in the stellar density profile found at more than 10 kpc from the galaxy centre. These peaks are likely due to misclassified satellites or mergers. This step helps isolate the stellar density profile from environmental noise, allowing for a more accurate measurement. In addition, we exclude the inner region within the simulation softening length from the 2D stellar density profile. After this post-processing, R$_1$ is determined by interpolating the remaining profile, to find where the 2D stellar surface density equals 1 M$_\odot$/pc$^{2}$. In addition, the stellar mass of each galaxy has been measured within 20\% of R$_{200}$.

Fig. \ref{fig:fig1} illustrates our methodology applied to a MW-like galaxy from each simulation suite. The left-hand panels show the 2D stellar density profile of each galaxy in purple, with circled markers indicating the profile after removing the softening length and potential artefacts. The vertical lines represent the stellar half-mass radius (solid green) and R$_1$ (dash-dot orange). In the right-hand renderings, we can appreciate the sizes of the stellar half-mass radius, R$_1$, as well as twice the stellar half-mass radius (indicated by a dashed green line). In each case, even twice the stellar half-mass radius captures only the inner part of the galaxy and does not encapsulate the full extent of the disk. In contrast, R$_1$ encloses most of the visible galaxy including the majority of the disk, in each simulation.

\begin{figure*}
    \centering    \includegraphics[width=\textwidth]{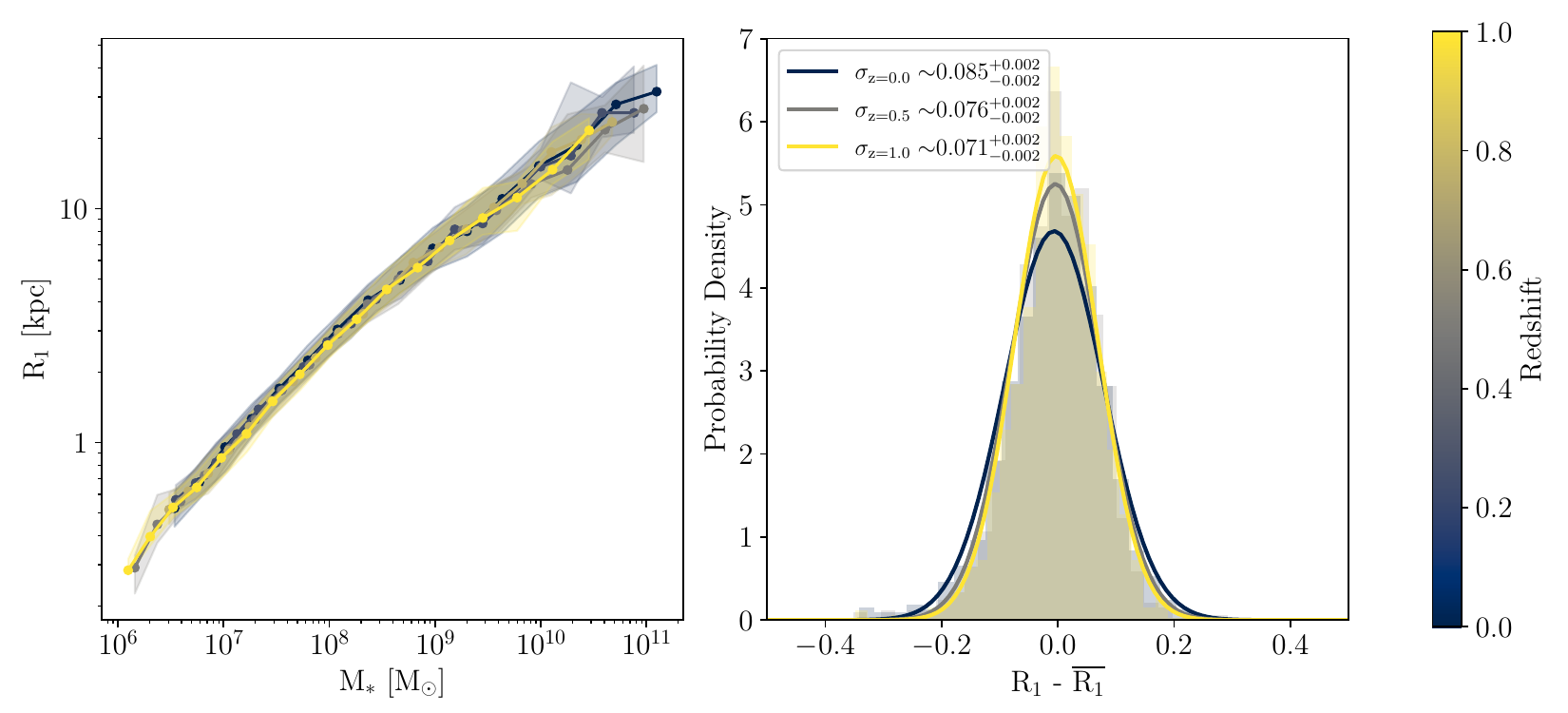}
    \caption{R$_1$ as a function of galaxy stellar mass over redshift for all the galaxies in our sample, where the stellar mass is measured within 20\% of R$_{\rm 200}$,. The left-hand panel shows the median value of the galaxy size, measured using the R$_1$ criteria, for all simulations. Each colour represents a different redshift, from $z$=0 (in blue) to $z$=1 (in yellow). The median value of the size has been measured within 20 stellar mass bins in each case. We also show the 1$\sigma$ error as shaded regions. The right-hand panel shows a normalized histogram of the scatter between each singular size value and its corresponding bin value, $\overline{\rm R}$, for $z$=0.0, $z$=0.5 and $z$=1 in blue, grey and yellow, respectively. In every case, the global $\sigma$ of the relation is found by fitting a Gaussian distribution (solid line following the same colour of the respective histogram). The respective $\sigma$ is defined as the FWHM of the respective Gaussian distribution whose errors are computed via bootstrapping.}

    \label{fig:redshift}
\end{figure*}

\begin{figure}
    \vspace{-0.3cm}
    \hspace{-0.7cm}   \includegraphics[width=1.2\columnwidth]{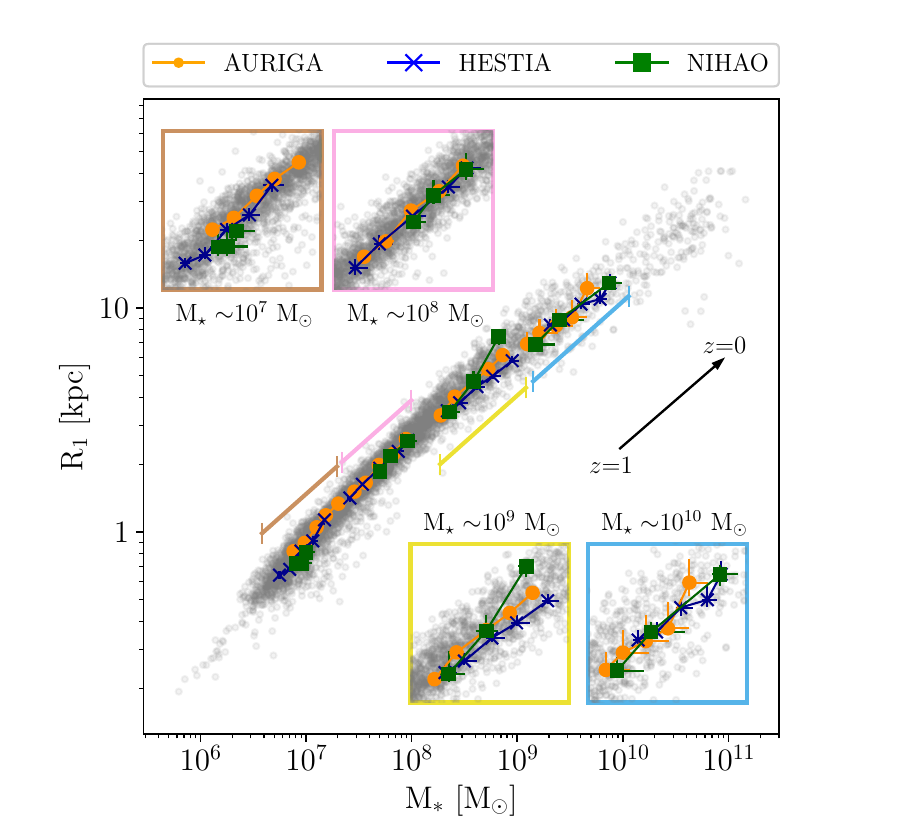}
    \caption{Evolution of R$_1$ as a function of galaxy stellar mass, measure within 20\% of R$_{\rm 200}$, over four different stellar mass ranges. For each simulation suite, we select $z$=0 galaxies at 10$^{7}$, 10$^{8}$, 10$^{9}$ and 10$^{10}$ M$_\odot$, within a bin of $\pm 0.5\rm dex$, and track their size-evolution over redshift up to $z$=1, using 5 increasing $z$, from left to right. Each of the stellar mass ranges selected are indicated and zoom-in with brown, pink, yellow and blue regions, respectively. The median value of the size and the stellar mass have been measured for each of the redshifts. Orange dots represent the size evolution of the \AURIGA suite, while blue crosses and green squares indicate the \HESTIA and \NIHAO samples, respectively. The whole sample at $z$=0, regardless of the simulation suite, is also shown as grey dots.}
    \label{fig:evolution}
\end{figure}

\begin{figure}[t!]
   \hspace{-0.2cm}
   \includegraphics[width=\columnwidth]{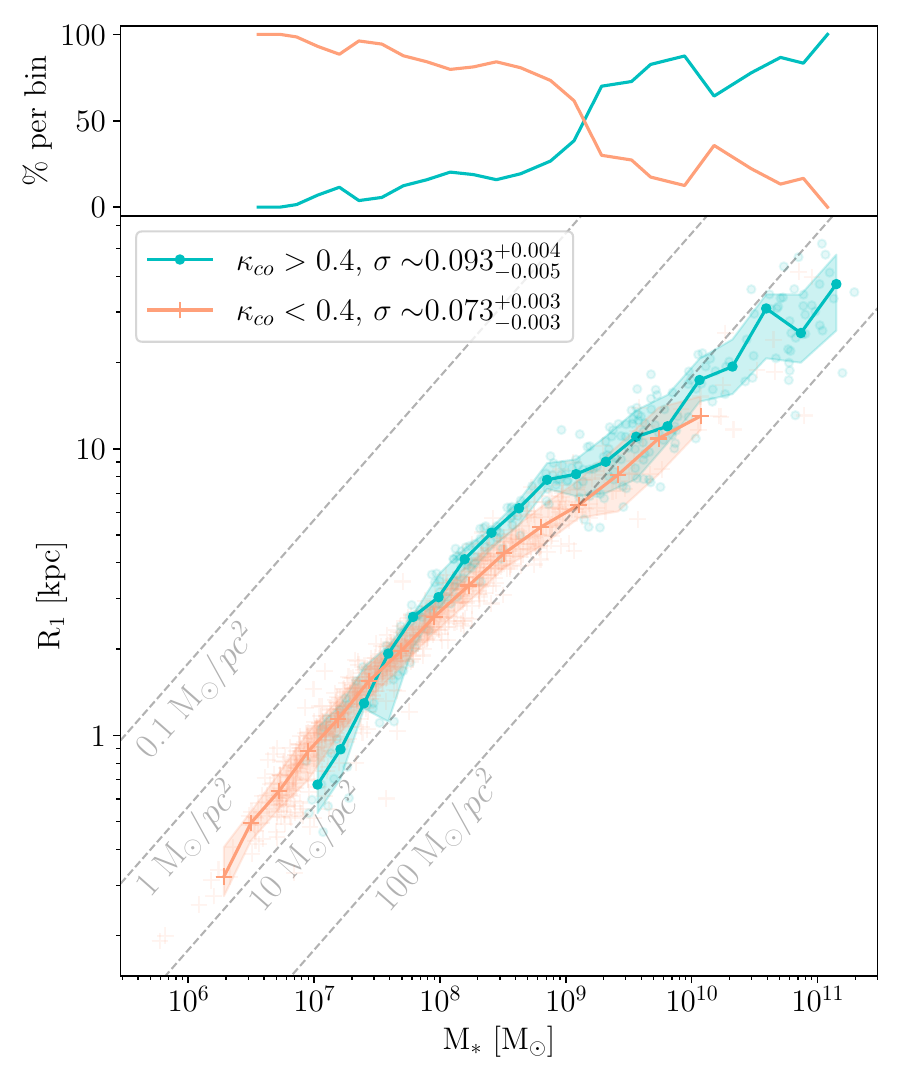}
   \caption{R$_1$ as a function of galaxy stellar mass, measure within 20\% of R$_{\rm 200}$, for all galaxies in the 3 simulations of our sample. Each of them is characterized as rotational (shown in blue) or dispersion (shown in pink) supported, following the criteria of \cite{Sales2010} and \cite{Correa2017}. The median size value has been measured across 20 stellar mass bins in each case. We also show as shadow regions the 1$\sigma$ error for both classifications. The top panel represent the percentage of galaxies in each morphology per bin, following the same colour-code as the main panel. The global FWHM of the relation is found by fitting a Gaussian distribution, and it is indicated in the legend as $\sigma$, whose errors are computed via bootstrapping.}
   \label{fig:morpho}
\end{figure}

\section{RESULTS}

\subsection{The R$_1$ - M$_{*}$ relation}
\label{sec:R1}

In this section, we present the R$_1$ - stellar mass relation for over 1000 galaxies with stellar masses between 4.9$\times$10$^5$ M$_\odot$ and 1.9$\times$10$^{11}$ M$_\odot$, using \AURIGA, \HESTIA and \NIHAO (Figs. \ref{fig:comparison} and \ref{fig:R1}). We compare our R$_1$ measurement with the stellar half-mass radius, R$_{1/2}$, i.e., we divide the stellar mass range into 20 stellar mass bins and compute the median and  1-sigma scatter of R$_1$ and R$_{1/2}$ for the galaxies falling into each bin.

The top panel of Fig \ref{fig:comparison} shows the size-stellar mass relation for the three different simulations used in this work, in which we plot the 1-sigma scatter of the median values as a shaded area. Over the entire stellar mass range, the relation differs when using R$_1$ compared to R$_{1/2}$ as a proxy for the size of galaxies. For the \AURIGA and \HESTIA samples, the R$_1$-M$_\star$ relation becomes steeper and larger than R$_{\rm 1/2}$-M$_\star$ for galaxies with stellar masses larger than $10^7$ M$_\odot$. In contrast, for \NIHAO, R$_{\rm 1/2}$ exceeds R$_1$ for M$_\star < 10^8$ M$_\odot$. This is because \NIHAO simulations produce "cored" dark matter haloes, which in turn give rise to ultra-diffuse galaxies, at M$_\star$$\sim$10$^8$ M$_\odot$, having therefore larger R$_{\rm 1/2}$ than \HESTIA or \AURIGA galaxies at the same stellar mass \citep{DiCintio2014a,DiCintio2014b,Tollet2016,DiCintio2017a}.

The bottom panel of Fig. \ref{fig:comparison} shows the scatter within the sample, being the R$_1$ scatter smaller than R$_{\rm 1/2}$ at every stellar mass bin. We compute this by measuring the difference between each size value and the corresponding bin value. The result can be fitted as a Gaussian distribution, where the center of the distribution aligns with the mean size value for each bin. In the legend of each plot, we indicate that the numerical values of the global scatter for R$_{\rm 1/2}$ is consistently larger than that of R$_1$, by factors of approximately $1.75$ (\AURIGA), $1.85$ (\HESTIA), and $1.34$ (\NIHAO), respectively. This is qualitatively consistent with the observation results of \cite{Trujillo2020} when comparing the scatter of the size-stellar mass using R$_1$ and R$_{\rm eff}$.

To assess the robustness of the R$_1$-stellar mass relation found in the simulations, we plot, in Fig. \ref{fig:R1}, the relation for the \AURIGA (orange), \HESTIA (blue) and \NIHAO (green) samples alongside one another. For comparison with observational data, we include dashed grey lines indicating constant projected stellar mass densities of 0.1, 1, 10 and 100 M$_\odot$/pc$^{2}$, from top to bottom, given by $S_\star = M_\star / \pi r^2$ 
\footnote{Note that in this paper we plot lines of constant $M_\star /\pi r^2$, which is a factor of $\pi$ different from the lines plotted in \cite{Trujillo2020}.}
. We also show, with a red dashed line, the corresponding values for \texttt{EAGLE} 
galaxies \citep{Claudio2025}, with stellar masses between $10^{7.5}$ M$_\odot$ and $10^{12}$ M$_\odot$. Additionally, in appendix \ref{FIRE} we extend our comparison with a sample of 8 low-mass galaxies from the public \texttt{FIRE-2} simulations (see Fig \ref{fig:R1-fire}). Note that \texttt{FIRE-2} galaxies were not included in the main work due to the limited number of galaxies available across the stellar mass range. However, it can be appreciated that both \texttt{EAGLE}
and \texttt{FIRE-2} follow the same trend as our simulation suites.

In agreement with \cite{Trujillo2020}, the median stellar density within R$_1$ lies above the 10 M$_\odot$/pc$^{2}$ line for galaxies with a stellar mass M$_\star < 5\times10^8$ M$_\odot$, heading towards higher stellar mass densities for the most massive galaxies with M$_\star > 10^9$ M$_\odot$. \cite{Chamba2024} suggests that feedback and environmental processes may regulate galaxy sizes more rapidly in the low mass regime, compared to their massive counterpart, leading to a break in the size-stellar mass relation at $\rm \sim$ 4$\times 10^{8} M_\odot$, where it crosses the 10 M$_\odot$/pc$^{2}$ line. We note that the transition occurs at different stellar masses in \NIHAO compared to \AURIGA and \HESTIA. This discrepancy may result from the different morphology of galaxies in each simulation sample: \AURIGA and \HESTIA massive galaxies tend to be disc-dominated spiral galaxies, while \NIHAO's massive galaxies predominantly exhibit elliptical morphology (see sec. \ref{sec:morpho} for more details). Observations suggest that the transition of R$_1$ at the 10 M$_\odot$/pc$^{2}$ line likely reflects differences in the formation and accretion history of the most massive galaxies compared to lower-mass galaxies. Different predominant morphology in each simulation could affect the stellar mass range in which this deviation occurs. All in all, galaxies with stellar masses from $10^7$ to $10^{11}$ M$_\odot$ follow a power law $\rm R_1 = M_\star ^{\beta}$ with $\beta$ = 0.375, consistent with observations that find similar behaviour with $\beta$ = 0.35 \citep{Trujillo2020} and $\beta$ = 0.377 \citep{Hall2012}. While different feedback schemes play a key role in defining the stellar half-mass radius \citep{Crain2015}, potentially explaining the discrepancies among simulations, R$_1$ remains independent of the specific simulation suite and feedback scheme chosen. This suggests that R$_1$ intrinsically captures the physics regulating star formation at the edges of galaxies.

\subsection{Redshift evolution}
\label{sec:redshift}
In the previous sections, we demonstrated that R$_1$ is a robust quantity for estimating galaxy size: it is consistent across the simulations suites probed. We now explore how R$_1$ and its associated scatter evolve with redshift for all simulation suites.

In the left panel of Fig. \ref{fig:redshift}, we present the median R$_1$ values in each stellar mass bin over different redshifts: $z$$\sim$$0.0$, $z$$\sim$$0.2$, $z$$\sim$$0.5$, $z$$\sim$$0.7$ and $z$$\sim$$1.0$. Each redshift is coloured following the colourbar on the right, with the 1-$\sigma$ scatter for each redshift shown as shaded regions. The evolution of the normalisation and slope of the mean size-stellar mass relation does not change notably over redshift, however, the scatter at fixed stellar mass appears to decrease with increasing redshift. The right-hand panel shows the distribution of sizes at $z$$\sim$1.0, $z$$\sim$0.5 and $z$$\sim$0.0, which quantifies the decreasing scatter with increasing redshift. This evolutionary trend in the scatter may be linked to the cumulative effect of merger activity. In other words, galaxies at higher redshifts have experienced fewer mergers, leading to less diversity in galaxy sizes. This result aligns with previous studies using R$_{\rm eff}$, such as \cite{Nedkova2021} or \cite{Kawinwanichakij2021}, which did not find significant changes in galaxy sizes across redshift. In contrast, \cite{Buitrago2024} found that, using R$_{\rm edge}$ as a proxy for the size, MW-like galaxies have doubled in size since $z$=1. Although \cite{Chamba2020} found no significant differences when using $\rm R_{edge}$ instead of $\rm R_1$, there is no guarantee that such a proxy based on MW-like galaxies should be able to characterize the radius to which in situ star formation takes places for all galaxies regardless of their evolution. However, due to the complexity of measuring R$_{\rm edge}$ for non-massive galaxies, we opted to use R$_1$. A more detailed study of size evolution across higher redshifts, based on a truncation in the stellar density profile, is deferred to future research.

In addition to studying the evolution of the size-stellar mass relation, simulations provide the opportunity to examine the evolution of the sizes of individual galaxies. A question arises: do the galaxies evolve along this relation? To address this, we selected galaxies at $z$=0 from each simulation sample in four stellar mass ranges within a bin of $\pm$0.5 dex: $10^7, 10^8, 10^9$ and $10^{10}$ M$_\odot$. We track these galaxies back in time up to $z$=1 \footnote{We select those galaxies that can be tracked up to $z$=1 by using the most massive progenitor at each snapshot defined by AHF’s merger tree.}, using 5 increasing $z$. Fig. \ref{fig:evolution} illustrates the median evolution for each stellar mass range and simulation suite. Generally, galaxies evolve along the size-stellar mass relation, regardless of the stellar mass range and simulation suite. An exception is found in the \NIHAO sample, where galaxies that lie in the size-stellar mass relation at $z$=1 for the stellar mass bin $\sim$10$^9$ M$_\odot$, ended with larger sizes at present-day redshift compared to galaxies in other simulation suites. Note that, even when present-day galaxies have larger sizes than the average, their progenitors still lie in the R$_1$ relation at earlier redshifts. The discrepancy in \NIHAO galaxies for this stellar mass range is explained by repeated SNe driven gas outflows, which build up larger galaxies with time, particularly in the stellar mass range M$_\star$$\sim$10$^{8-9}$ M$_\odot$ \citep{DiCintio2014a,DiCintio2014b,DiCintio2017a,Pontzen2012}. 
The initial selection and morphology of \NIHAO galaxies (see Fig. \ref{fig:R1}) indicate that they tend to be larger than those in \AURIGA and \HESTIA samples for stellar masses between 10$^8$ and 10$^9$ M$_\odot$. Another exception is the evolution in the low mass range. While the stellar mass in \AURIGA and \HESTIA galaxies seem to evolve continuously and monotonically along the relation over the four stellar mass ranges, the \NIHAO galaxies exhibit a different evolution for dwarfs: galaxies with 10$^7$ M$_\odot$ stellar masses evolve very little in stellar mass or size, and galaxies in the $\sim$10$^8$ M$_\odot$ stellar mass range at $z$=0 are larger at higher redshifts compared to those from \AURIGA and \HESTIA. Different feedback schemes could be the reason for evolution discrepancies between the simulation samples. This may be particularly true in the low mass regime, where the modelling and implementation of a relatively low number of feedback events in relatively shallow dark matter halo potentials can significantly shape galaxy properties. Overall, we demonstrated that galaxies evolve across the R$_1$-M$_\star$ relation at various $z$, despite their specific stellar mass or simulation used.

\subsection{Morphology dependence}
\label{sec:morpho}

\begin{figure}[t!]
   \hspace{-1.5cm}
   \includegraphics[width=1.3\columnwidth]{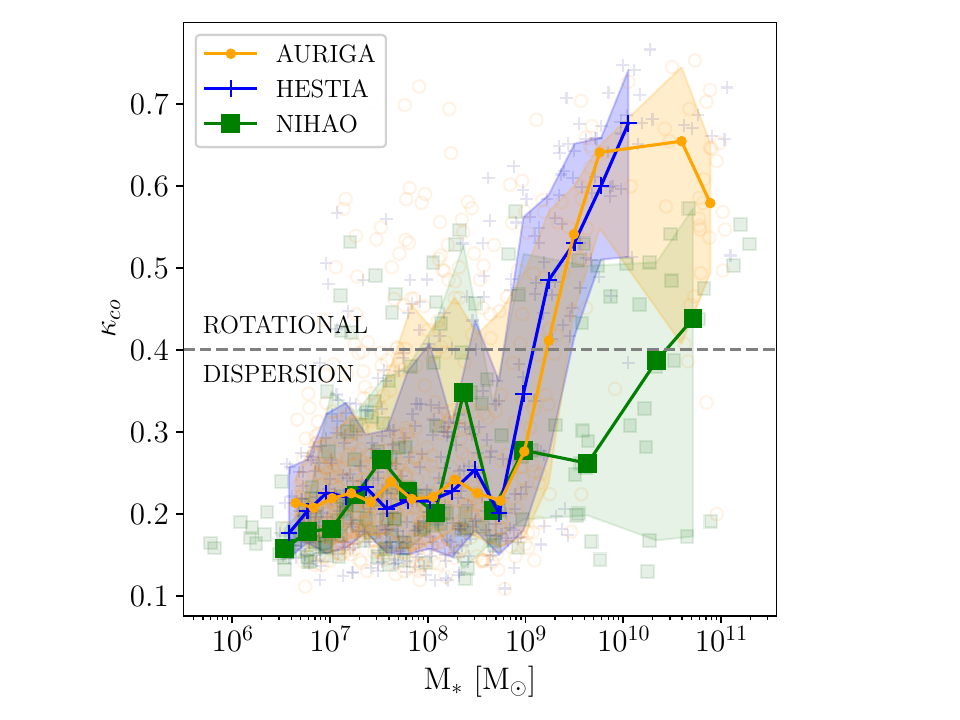}
   \caption{Rotational velocity versus stellar mass for each simulation suite. We show the median value of the size measure within 20 stellar mass bins. Orange, blue and green relations correspond to the \AURIGA, \HESTIA and \NIHAO samples. We also show as orange, blue and green shadow regions, the 1$\sigma$ for \AURIGA, \HESTIA and \NIHAO, respectively. Galaxies above the horizontal dashed line are considered rotational-supported and viceversa (see \cite{Sales2010} and \cite{Correa2017} for more details about the selection criteria).}
   \label{fig:kcosims}
\end{figure}

We showed that using R$_1$ to measure the global size-stellar mass relation of galaxies decreases the scatter by about twice compared to using R$_{\rm 1/2}$. We now check whether this is also the case for different morphologies. We examined the R$_1$ - stellar mass relation by categorizing sample galaxies into two groups, regardless of the simulation used: rotational supported and non-rotational, i.e., dispersional supported. We classified galaxies as rotational supported following the criteria from \cite{Correa2017}, where galaxies with $\kappa_{co}>0.4$, are distinguished as spirals and those with $\kappa_{co}<0.4$ as elliptical. Here, $\kappa_{co}$ is the fraction of the kinetic energy of the galaxy invested in ordered corotation (see \cite{Sales2010} and \cite{Correa2017} for more details). Fig. \ref{fig:morpho} illustrates the correlation between larger sizes and higher co-rotational velocities, showing that spiral galaxies tend to have larger R$_1$ sizes than elliptical galaxies at fixed stellar mass, for a stellar mass larger than $\sim$4$\times$$10^7$ M$_\odot$. This finding agrees with observational results from \cite{Trujillo2020}. Below this stellar mass, we can appreciate a change in the behaviour, in which the trend reverses. Dispersion-supported galaxies host larger sizes than rotational galaxies at lower stellar masses. However, this could be due to statistical limitations -since the sample of rotational galaxies at those stellar masses may not be representative in this case (see top panel of Fig. \ref{fig:morpho})- dispersion-dominated dwarfs tend to be cored galaxies and might have larger sizes than the more disky ones. It is important to note that in our sample, elliptical and spiral galaxies coexist within the same stellar mass range. As discussed in section \ref{sec:R1}, observed galaxies from \cite{Trujillo2020} reach higher stellar masses, where the distinction between ellipticals and spirals becomes more pronounced, with limited overlap between the two morphologies. While we cannot replicate galaxies in this most massive region, we still observe differences in the slope between elliptical and spiral morphologies, consistent with their findings in the stellar mass range where both morphologies coexist. 

For both cases, we have measured the 1-$\sigma$ scatter, finding a smaller scatter, $\sigma$$\sim$0.073$\pm$0.003, in those galaxies supported by dispersion. This result may seem in contradiction with previous discussions, primarily since R$_1$ was designed for MW-like galaxies. However, the scatter has been measured across the entire stellar mass range—from $10^7$ M$_\odot$ to $10^{11}$ M$_\odot$ for rotation-supported galaxies and $10^6$-$10^{10}$ M$_\odot$ for dispersion-supported galaxies. Thus, the scatter found for disk galaxies cannot be interpreted as the general behaviour of MW-like galaxies.

Fig. \ref{fig:kcosims} shows the general co-rotational behaviour concerning the galaxy's stellar mass for each simulation suite. In the low-mass regime, all galaxies, regardless of the simulation, tend to exhibit elliptical morphologies, with the median co-rotational value remaining below 0.4, until the galaxies reach a stellar mass of about 10$^9$ M$_\odot$. For more massive galaxies, we can appreciate a large difference between \AURIGA and \HESTIA galaxies and \NIHAO galaxies. The reason for this difference can be ascribed to the different selection criteria originally adopted in constructing the three simulated samples. While \HESTIA and \AURIGA galaxies are popularly used for the study of MW-like galaxies and their environment, \NIHAO galaxies do not target any particular morphological type. However, \NIHAO tends to harbour a high population of elliptical galaxies. An exception to this trend occurs in those galaxies with a stellar mass range between $10^{8}$ and $10^{9}$ $\rm M_\odot$, where the \NIHAO sample has a greater number of spirals than their counterparts in the other simulations. This behaviour is also reflected in Figs. \ref{fig:R1} and \ref{fig:evolution}, where higher values of R$_1$ are found for \NIHAO galaxies within this stellar mass regime.

 \subsection{The connection with halo properties} \label{sec:halo}

\begin{figure}[t!]%
    \hspace{-0.2cm}{{\includegraphics[clip,trim={1.2cm 0 0 1cm},width=1.25\columnwidth]{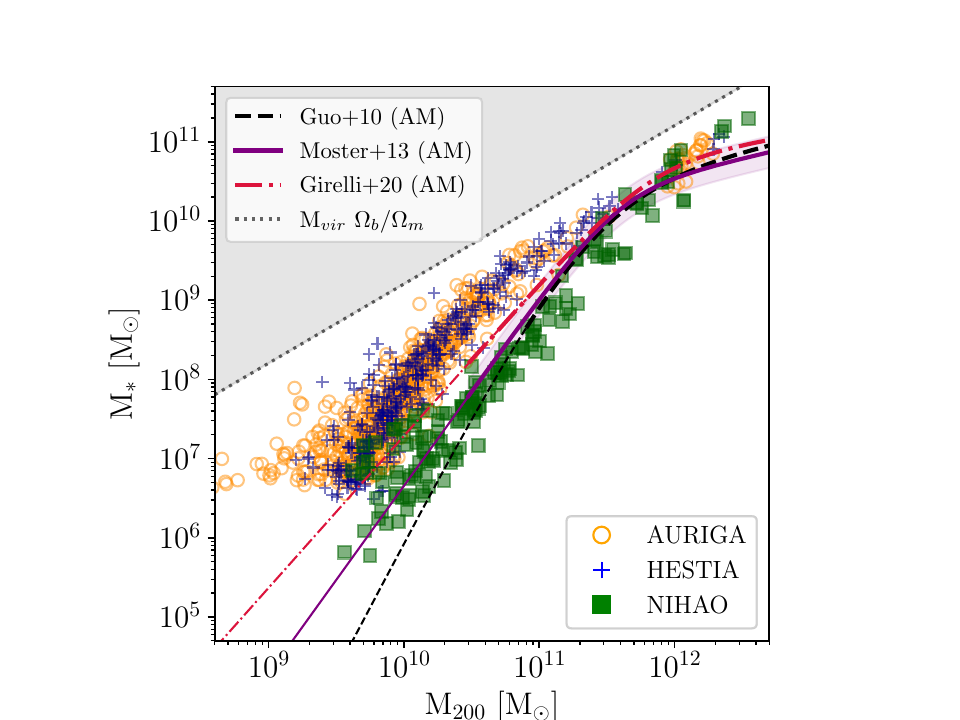} }}
    
    \hspace{-0.2cm}
{{\includegraphics[clip,trim={1.2cm 0 0 1cm},width=1.25\columnwidth]{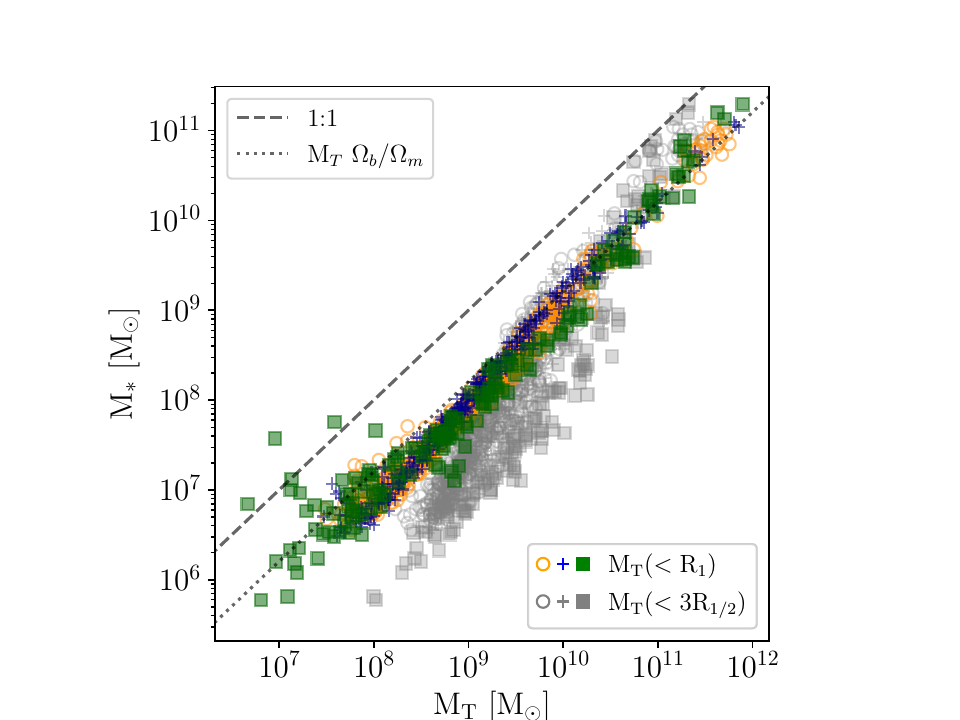} }}
    
    \caption{Upper panel: stellar-to-halo mass relation (SHMR) for each of the simulation suites. In solid purple, the abundance matching relation from \cite{Moster2013} is shown with a 0.2 dex scatter. The dashed dark and dot-dashed magenta lines represent estimates by \cite{Guo2010} and \cite{Girelli2020}, respectively. Thick lines correspond to the range where the abundance matching relations are constrained, while the thin lines are extrapolation. Bottom panel: Stellar mass versus total mass relation within R$_1$ (STMR), where the stellar mass is measured within 20\% of R$_{\rm 200}$. The dashed line represents the 1:1 relation. The universal baryon fraction is shown with a dotted line in both cases. Empty orange dots, blue crosses and green squares represent the \AURIGA, \HESTIA and \NIHAO, respectively. For comparison, we also show in grey squares, dots and crosses the STMR within 3R$_{\rm 1/2}$.}

    \label{fig:HALO}
\end{figure}

In the previous section, we showed that the scatter in R$_1$ is generally a factor of almost 2 smaller than that found in R$_{\rm 1/2}$. We now explore R$_1$ as an alternative metric to R$_{\rm 1/2}$ when investigating the connection between galaxies and their host dark matter halos.

The scatter of the SHMR is an interesting quantity because it reflects the efficiency of galaxy formation at a fixed halo mass: galaxies with higher stellar masses have been more efficient in forming stars than their less massive counterparts. Thus, this scatter is expected to be linked to the baryonic cycle of galaxies; however, how it connects to galaxy properties remains a matter of debate \citep{Wechsler2018}.
With halo mass estimations being one of the main caveats to explore this issue observationally, \cite{Scholz2024} introduce the stellar-to-total dynamical mass relation (STMR) as a direct observational alternative for studying the galaxy-halo connection, with total dynamical masses measured within 3R$_{\rm eff}$. They find that galaxy properties exhibit similar trends across the scatter of both the SHMR and the STMR, with total dynamical masses correlating with halo masses at a fixed stellar mass. Based on this, they argue that total dynamical mass measured within 3R$_{\rm eff}$ serves as a proxy for halo mass and the scatter of the STMR can be used as an alternative to that of the SHMR. In this section, we investigate whether total masses measured within R$_1$ are able to trace their halo masses. If so, R$_1$ could serve not only as a reliable size measure but also as a direct method for studying the connection between galaxies and their halo properties. Analogous to \cite{Claudio2025} work — who explore the R$_1$-R$_{200}$ for \texttt{EAGLE} galaxies, finding a notable small scatter that can be used to infer properties of their host halos — we measure the total mass as contained within R$_1$, and later compare with the observational results from \cite{Scholz2024}, who computed the total dynamical mass following 3D Jeans Equations (see \cite{Zhu2023} for more details).

Fig \ref{fig:HALO} shows both SHMR (top panel) and the STMR inside R$_1$ and 3R$_{\rm 1/2}$ (bottom panel) using coloured and grey markers, analogous to the STMR measured within 3R$_{\rm eff}$ from \cite{Scholz2024}. Note that R${\rm eff}$ in observational studies is derived from a photometric band that is not always available in our sample. For this reason, we use R$_{1/2}$ for comparison, as it provides a more consistent measurement across different simulations. In the top panel, the purple, dark and magenta curves represent semi-empirical relations from \cite{Moster2013}, \cite{Guo2010} and \cite{Girelli2020} respectively. While \NIHAO galaxies follow the \cite{Moster2013} semi-empirical relation, \AURIGA and \HESTIA are placed above this relation for the M$_{200} < 10^{11}$ M$_\odot$ range. However, note that below this halo mass, a large uncertainty exists in the predicted relation (see \cite{Girelli2020} for more details). Whether a simulation matches the abundance matching relations depends significantly on how feedback mechanisms are implemented. For example, while \NIHAO is calibrated to match the abundance matching relation across their high mass ranges (see Fig 8. of \cite{Schaye2015}), the \NIHAO sample used in this paper does not include AGN feedback ( see \cite{Blank2019} for AGN implementation in \NIHAO) and yet reproduces the SHMR across a large galaxy mass range. In contrast, \AURIGA and \HESTIA, though not explicitly calibrated to match the abundance matching relation, incorporate both AGN and SN feedback.


Given the robustness of the R$_1$-stellar mass relation across the different simulation suites demonstrated in the previous section, we may expect a stronger correlation between the total mass within R$_1$ and the stellar mass (STMR) compared to the SHMR. The bottom panel of Fig. \ref{fig:HALO} shows the former for each simulation suite, with notably reduced scatter at the low-mass end compared to the SHMR. Additionally, the STMR within 3R$_{\rm 1/2}$ is shown in grey. For both STMR cases, the scatter for each simulation suite has been measured (not shown), revealing a systematic reduction when using R$_1$. The median scatter of STMR within R$_1$ is found to be $\sigma_{\rm R1}$$\sim$0.15$\pm$0.01, whereas the scatter measured within 3R$_{\rm 1/2}$ is $\sigma_{\rm 3R1/2}$$\sim$ 0.28$\pm$0.02.
Alongside the STMR relation, we also show the 1:1 line and the universal baryon fraction with dashed and dotted lines, respectively. The fact that galaxies closely follow the universal baryon fraction line can be explained by R$_1$ corresponding to the radius where the gas density required for SF reaches a minimum. Under this assumption, all stellar mass is enclosed within R$_1$. Galaxies with stellar mass below 10$^9$ M$_\odot$ fall below the universal baryon fraction relation, while MW-like galaxies seem to match it. We found several outliers below 10$^8$ M$_\odot$ for the \NIHAO sample, with stellar masses larger than expected for a fixed total mass, even surpassing the 1:1 line. For galaxies above this line, we inspected the stellar surface density profile (not shown here). In each of these galaxies, the maximum values in their stellar density barely reached the $\Sigma_\star$=1M$_\odot$/pc$^2$ threshold, leading to R$_1$ values not capable of enclosing the whole galaxy. This threshold is defined for MW-like galaxies and differs for dwarfs or elliptical galaxies (see \cite{Trujillo2020} and references therein for more details), thus, low-mass galaxies whose sizes cannot be explained by using R$_1$ are to be expected. However, note that for galaxies where the maximum value of the stellar surface density profile is well above the density threshold, we can still recover the expected size-stellar mass relation. 

For a visual inspection, Fig \ref{fig:HALO-fire} extend our results with 8 low-mass \texttt{FIRE-2} galaxies, which lie in the same space as other simulation suites. In addition, we present in appendix \ref{apen-cum} the total cumulative mass versus radius for four different stellar masses: $\sim$10$^{10}$, 10$^9$, 10$^8$ and 10$^7$ M$_\odot$. While the cumulative total mass distribution evolves differently across the entire radius depending on the simulation suite used, the median profiles of each sample consistently intersect at $\sim$R$_1$, regardless of the stellar mass range. This implies that \AURIGA,  \HESTIA and \NIHAO galaxies tend to have the same amount of total mass at the R$_1$ range (shown in grey), diverging again at larger radii. Galaxies from the \texttt{FIRE-2} sample that fall within the corresponding stellar mass ranges are also shown, intersecting with the other samples at R$1$ as well. However, since only a few \texttt{FIRE-2} galaxies fall into this range, we may not be able to interpret them as a systematic trend.

Although the focus of this work is not on the baryonic processes that influence the scatter in the SHMR, examining the halo mass derived from the total mass within R$_1$ could provide valuable insights into the relationship between these processes. To assess whether the total mass within R$_1$ accurately traces the halo mass — and thus whether the scatter in the STMR could serve as an alternative to the SHMR — we conduct an analysis similar to that of \cite{Scholz2024} in Appendix \ref{appen-laura}. It is important to note, however, that we assume R$_1$ fully encompasses the stellar mass of galaxies, independent of their morphology. A more detailed exploration of the baryonic processes influencing halo properties will be addressed in future work.

\section{CONCLUSIONS}\label{sec:conclusions}

The sizes of galaxies have been a subject of extensive study over the past few decades, as they are essential in our understanding of how galaxies form and evolve. However, it is only in recent years that observational studies have begun to explore the borders of galaxies using physically motivated definitions. Specifically, these studies have employed the location of gas density threshold for star formation as a natural size indicator \citep{Trujillo2020,Chamba2022,Buitrago2024}. The radius R$_1$, which serves as a proxy for this critical gas density threshold \citep{Martinez-Lombilla2019}, has been shown observationally to reduce the scatter in the size-stellar mass relation of galaxies. In this paper, we aim to assess the reliability of this new definition across different state-of-the-art hydrodynamical simulation suites. The key findings of our work are summarised as follows:

\begin{itemize}

\item The size–stellar mass relation obtained using R$_1$ shows a systematic reduction in scatter by a factor of $\sim$2 dex compared to the relation defined by R$_{\rm 1/2}$, for all simulation suites (Figs \ref{fig:comparison} and \ref{fig:R1}). These results are in agreement with observations of \cite{Trujillo2020}, including the general trend, where massive galaxies above 10$^{9}$ M$_\odot$ begin to exhibit a shallower size-stellar mass slope, as shown in Fig \ref{fig:R1}.

\item We study the evolution of the size-stellar mass relation up to $z$$\sim$1 and find that the slope and normalisation of the relation are redshift-independent, and that the scatter decreases with increasing redshift (Fig. \ref{fig:redshift}). We found that most simulated galaxies have evolutionary tracks that lie along the median $\rm R_1$-stellar mass relation.

\item Regarding morphology, we divided the whole sample into rotational and dispersion-supported galaxies, based on the criteria of \cite{Correa2017} (Fig. \ref{fig:morpho}). For stellar masses above 10$^{7}$ M$_\odot$, rotationally-supported galaxies generally exhibit larger sizes compared to those that are dispersion-supported, indicating that the new size definition R$_1$ is correlated with the morphology of galaxies. Below 10$^7$ M$_\odot$, we find larger sizes in galaxies supported by dispersion. However, it is important to note that, in this lower mass range, the number of rotationally-supported galaxies is low.

\item Although each simulation suite is strongly influenced by the feedback models implemented, the use of R$_1$, not only as a measure of galaxy size but also as a means to study the connection with halo properties, yields a stronger correlation between stellar mass and total mass enclosed within R$_1$ compared to the traditionally used SHMR (Fig. \ref{fig:HALO}). For galaxies with stellar masses above 10$^9$ M$_\odot$, we observe a consistent correlation in the scatter between SHMR and STMR (Fig. \ref{fig:delta}), being also consistent with properties that correlate with the SHMR scatter, similar to the findings of \cite{Scholz2024} using 3R$_{\rm eff}$. This suggests that R$_1$ could also serve as a viable observational tracer of the halo mass. Those baryonic processes that influence the galaxy at fixed stellar mass, shaping the scatter of SHMR, can also be inferred by studying the scatter in the STMR. 

\end{itemize}

We caution the reader that the stellar and halo mass ranges explored in this work are upper limited by Milky-Way size galaxies, and do not cover the most massive range. In the future, extending the sample to include larger galaxies will provide valuable insights into identifying the slope change between elliptical galaxies and MW-like, discussed throughout the paper. Additionally, while we used R$_1$ as a size indicator instead of R$_{\rm edge}$, it is important to note that although both quantities are observed to be very similar at $z$$\sim$0, their differences may become significant at higher redshifts. A thorough investigation of R$_{\rm edge}$ will be essential for understanding changes in galaxy sizes and their evolution over time.

\section*{Acknowledgements}

 The authors sincerely thank Dr. Ignacio Trujillo and Dr. Laura V. Sales for their valuable insights. E. Arjona-Gálvez acknowledges support from the Agencia Espacial de Investigación del Ministerio de Ciencia e Innovación (\textsc{AEI-MICIN}) and the European Social Fund (\textsc{ESF+}) through a FPI grant PRE2020-096361. SCB  acknowledges financial support from the Spanish Ministry of Science and Innovation (MICINN) for PID2021-128131NB-I00 (CoBEARD) and CNS2022-135482 projects. RG acknowledges financial support from an STFC Ernest Rutherford Fellowship (ST/W003643/1). 
A. Di Cintio thanks the University of La Laguna (ULL) for financial support through a “Noveles Investigadores” grant, and the Ministry of Science and Innovation for additional funding from the EU Next Generation Funds under the 2023 Consolidación Investigadora program (grant code CNS2023-144669). CDV also acknowledges support from MICIU in the early stages of development of this work through grants RYC-2015-18078 and PGC2018-094975-C22. JAB is grateful for partial financial support from NSF-CAREER-1945310 and NSF-AST-2107993 grants and data storage resources of the HPCC, which were funded by grants from NSF (MRI-2215705, MRI-1429826) and NIH (1S10OD016290-01A1). NIL acknowledges funding by the European Union’s Horizon Europe research and innovation program (EXCOSM, grant No. 101159513). AK is supported by the Ministerio de Ciencia e Innovaci\'{o}n (MICINN) under research grant PID2021-122603NB-C21 and further thanks Teenage Fanclub for thirteen.

This research made use of the \textit{LaPalma} HPC cluster, under project \textit{can43}, PI A. Di Cintio, as well as the High-Performance Computers located at the Instituto de Astrofisica
de Canarias. The authors thankfully acknowledge the technical
expertise and assistance provided by the IAC (Servicios Informáticos Específicos, SIE) and the Spanish Supercomputing
Network (Red Española de Supercomputacion, RES).

\bibliographystyle{aa}
\bibliography{R1sizematters}

\appendix

\section{Comparison with FIRE-2 simulations}\label{FIRE}

\begin{figure}[h!]
\hspace{-1.5cm}
\includegraphics[width=1.4\columnwidth]{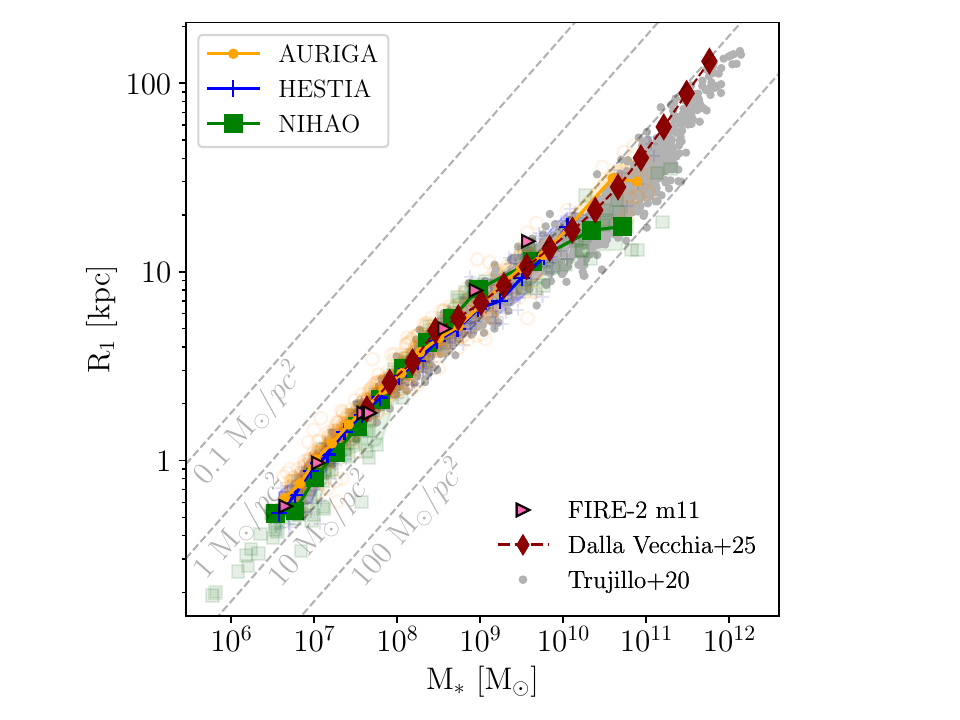}
   \caption{R$_1$ as a function of the stellar mass of galaxies, measured within 20\% of R$_{\rm 200}$, for different simulation suites as indicated in the legend. We show the median value of R$_1$ measured within 20 stellar mass bins, equally spaced. Orange, blue and green relations correspond to the \AURIGA, \HESTIA and \NIHAO sample. In addition, we also show in pink single \texttt{FIRE-2} low-mass galaxies and the relation found by \cite{Claudio2025}, using public data from \texttt{EAGLE}
   simulations in red. To compare with \cite{Trujillo2020}, we add in grey their observational results as well as dashed lines, corresponding to locations in the place with a constant projected stellar mass density of (from top to bottom) 0.1, 1, 10 and 100 M$_\odot$/pc$^{2}$.}
   
   \label{fig:R1-fire}
\end{figure}

\begin{figure}[t!]
    \hspace{-0.2cm}{{\includegraphics[clip,trim={1.2cm 0 0 1cm},width=1.25\columnwidth]{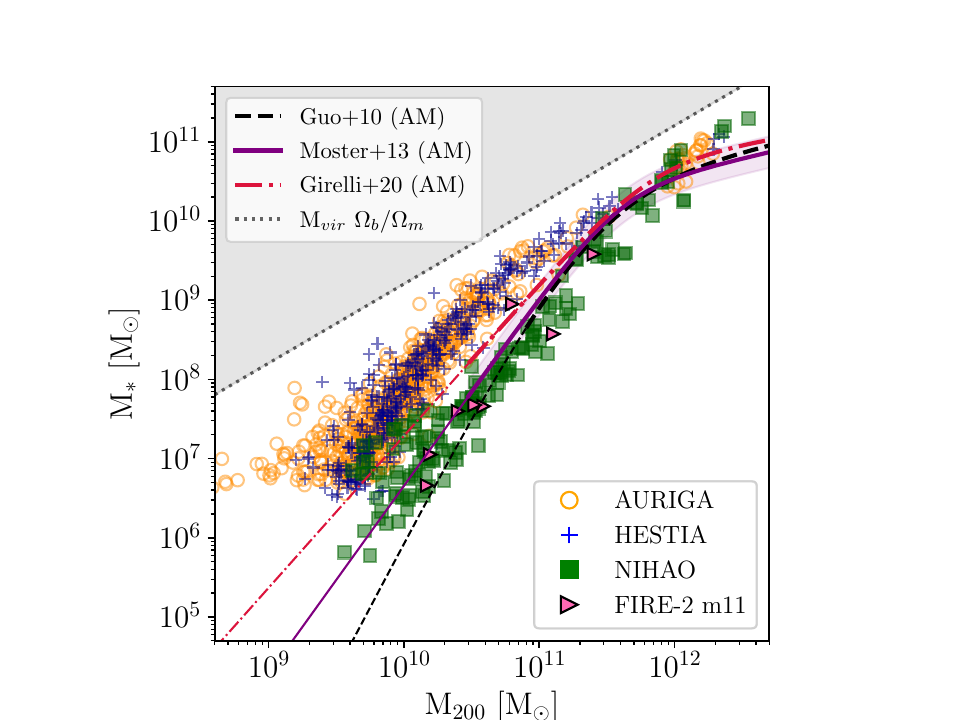} }}
    
    \hspace{-0.2cm}
{{\includegraphics[clip,trim={1.2cm 0 0 1cm},width=1.25\columnwidth]{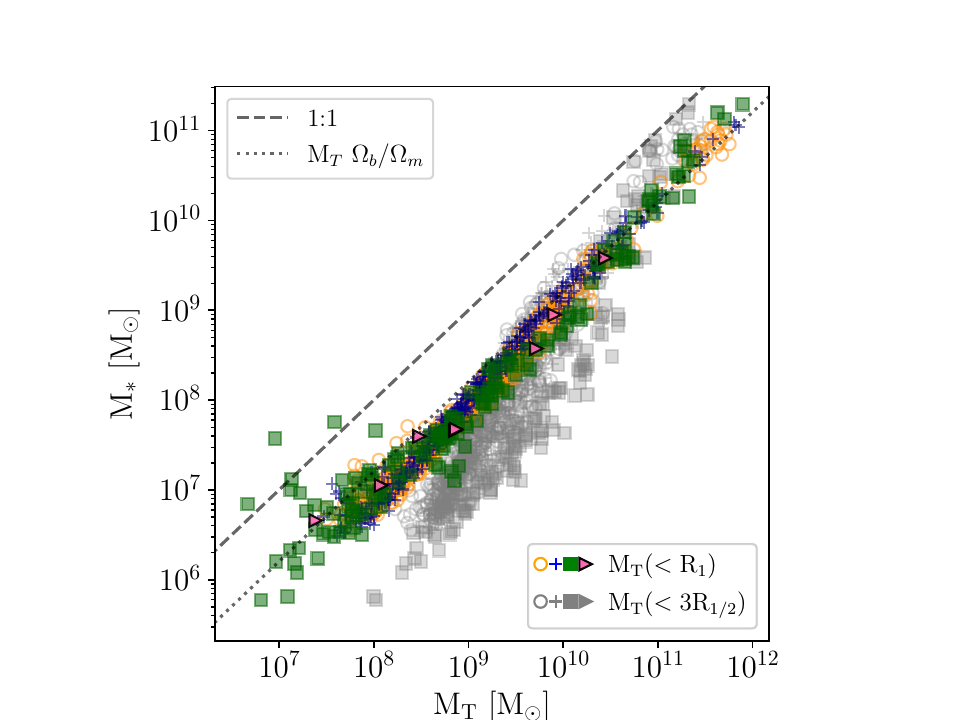} }}
    
    \caption{Upper panel: stellar-to-halo mass relation (SHMR) for each of the simulation suites. In solid purple, the abundance matching relation from \cite{Moster2013} is shown with a 0.2 dex scatter. The dashed dark and dot-dashed magenta lines represent estimates by \cite{Guo2010} and \cite{Girelli2020}, respectively. Thick lines correspond to the range where the abundance matching relations are constrained, while the thin lines are extrapolation. Bottom panel: Stellar mass versus total mass within R$_1$ (STMR), where the stellar mass is measured within 20\% of R$_{\rm 200}$. The dashed line represents the 1:1 relation. The universal baryon fraction is shown with a dotted line in both cases. Empty orange dots, blue crosses, green squares and pink triangles represent the \AURIGA, \HESTIA, \NIHAO and \texttt{FIRE-2} samples, respectively. For comparison, we also show in grey squares, dots and crosses the STMR within 3R$_{\rm 1/2}$.}

    \label{fig:HALO-fire}
\end{figure}

In this appendix, we extend Figs. \ref{fig:R1} and \ref{fig:HALO} in the respective Figs \ref{fig:R1-fire} and \ref{fig:HALO-fire} by adding a sample of 8 low-mass galaxies m11's runs \citep{Chan2018, El-Badry2018, Hopkins2018} from the public suite of \texttt{FIRE-2} zoom-in simulations (see \cite{Wetzel2023} for more details about \texttt{FIRE-2} cosmological simulations). We follow the same selection criteria described in section \ref{sec:method}. Note that we do not include \texttt{FIRE-2} galaxies in the main study due to the limited number of galaxies available across the stellar mass range. For this reason, these galaxies were included only for visual comparison.

\clearpage

\section{Cumulative total mass profiles}\label{apen-cum}

\noindent\begin{minipage}{\textwidth}
\includegraphics[width=\textwidth]{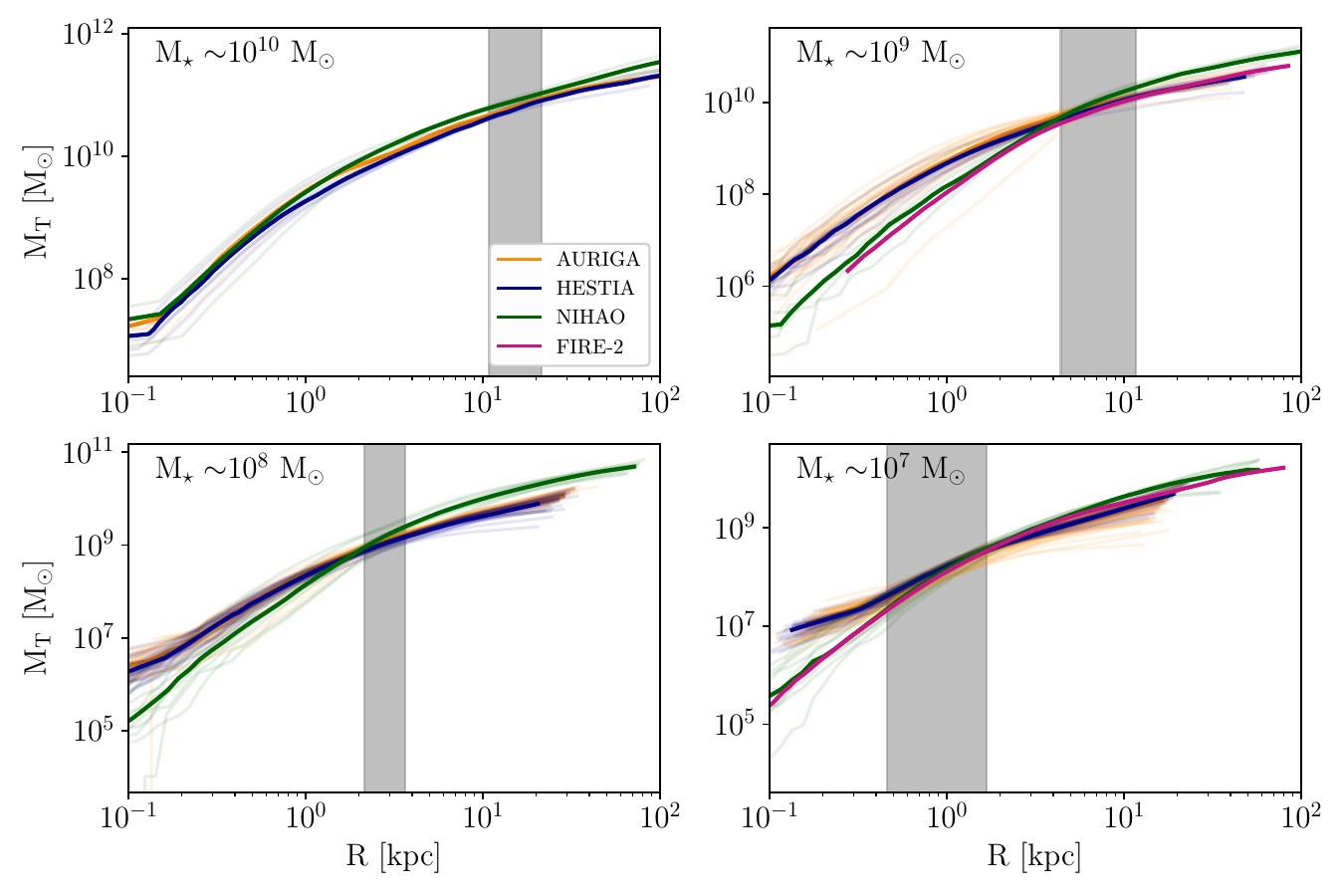}
\captionof{figure}{ Cumulative total mass versus radius across four stellar mass ranges. Each simulation suite is coloured following the colour code regarding the simulation suite. For each simulation suite, we select $z$=0
galaxies at 10$^{10}$, 10$^9$, 10$^8$ and 10$^{7}$ M$_\odot$ (from top-left to bottom-right), within a bin of ±0.5dex. The median cumulative total mass profile is shown with a solid thick line in orange, blue, green and pink for the \AURIGA, \HESTIA, \NIHAO and \texttt{FIRE-2} galaxies, respectively. The vertical grey region covers all the R$_1$ values found for each stellar mass range.}
\end{minipage}

\clearpage

\section{Total mass versus halo mass}\label{appen-laura}

\begin{figure}[h!]
\vspace{0.3cm}
\hspace{-0.7cm}
\includegraphics[width=1.2\columnwidth]{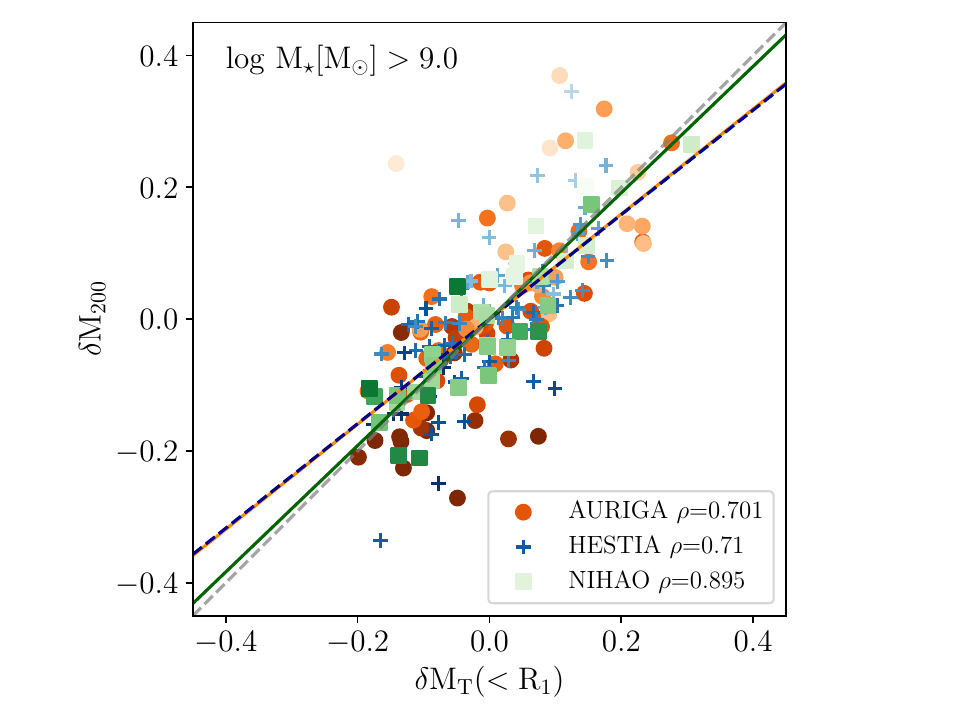}
   \caption{Halo mass versus total mass within R$_1$. We show the $\rm \delta M_{200}$ and $\rm \delta M_{T}(< R_1)$ residuals from the $\rm \log M_{200}$ $-$ $\rm \log M_\star$ versus $\rm \log M_T(< R_1)$ $-$ $\rm \log M_\star$ relations, for galaxies with $\rm M_\star > 10^9 M_\odot$. \AURIGA, \HESTIA and \NIHAO galaxies are shown with orange dots, blue crosses and green squares. Every galaxy is coloured by the concentration of its halo, by darker colour referring to higher concentration and viceversa. Every suite is upper and lower limit by an halo concentration value of 40 and 7, respectively. Best-fitting is shown following the same colour code used for each simulation suite, along with their Spearman rank correlation coefficients. We also show within dashed grey the 1:1 line for reference.}
   \label{fig:delta}
\end{figure}

We conduct an analogous analysis from \cite{Scholz2024} (see their Fig. Extended Data Fig. 2 and Methods section) to investigate whether the scatter of the STMR can be used as an alternative metric to the SHMR. Fig. \ref{fig:delta} reflects the residuals of $\rm M_{200}$ and $\rm M_T$ for galaxies above M$_\star$ > 10$^9$ M$_\odot$. We choose this stellar mass range to exclude low-mass dwarf galaxies for which the scatter in stellar mass (at fixed $\rm M_{200}$) is heavily influenced by baryonic physics.
The scatter has been measured by assuming both SHMR and STMR for galaxies above a stellar mass of 10$^9$ M$_\odot$ follow a linear distribution.

The positive correlation indicated by the Spearman rank coefficients means that, at fixed stellar mass, galaxies with higher total masses within R$_1$, $\rm M_T$, tend to have larger $\rm M_{200}$. To highlight the importance of the correlation between STMR's and SHMR's scatters, we have coloured every galaxy by its halo concentration\footnote{We use the halo concentration values from \texttt{AHF}, which numerically computes the concentration using Eq. 9 from \cite{Prada2012}.}, one of the properties that correlate with the SHMR scatter \citep{Matthee2017,Zu2021}. Dark colours refer to higher concentration values and vice versa. At fixed stellar mass, halos with positive scatter values, i.e. larger halo/total masses, have a lower concentration than those of smaller halo/total masses. Such behaviour indicates that those properties responsible for the scatter in the SHMR could be inferred by inspecting their scatter in STMR. This result aligns with the observational findings of \cite{Scholz2024}, who used the total dynamical mass within 3R$_{\rm eff}$ to investigate several baryonic processes that affect the STMR scatter.

\end{document}